\documentclass[twocolumn,twocolappendix]{aastex631}
\usepackage{float}

\usepackage{hhline}

\usepackage{xcolor}

\usepackage{appendix}

\begin{document}

%\title{A Wide Diversity of Dust Reservoirs of Lensed Quiescent Galaxies Reveal Both Rapid Destruction and Regrowth}
\title{Dust Survival in Lensed Quiescent Galaxies at $z>1$ Depends on How They Quench}
%Dust Reservoirs in Lensed Quiescent Galaxies Reveal Both Rapid Destruction and Regrowth
%Alternative title: Some Like It Harsh: Massive Dust Reservoirs in Lensed Quiescent Galaxies Illuminate Different Modes of Dust Destruction and Quenching Across Cosmic Time

\correspondingauthor{Sarah Bodansky}
\email{sbodansky@umass.edu}

\author[0000-0001-8246-1676]{Sarah Bodansky}
\affiliation{Department of Astronomy, University of Massachusetts, Amherst, MA 01003, USA}

\author[0000-0001-7160-3632]{Katherine E. Whitaker}
\affiliation{Department of Astronomy, University of Massachusetts, Amherst, MA 01003, USA}
\affiliation{Cosmic Dawn Center (DAWN), Denmark}

\author[0000-0002-3475-7648]{Gourav Khullar}
\affiliation{Department of Astronomy \& the DiRAC Institute, University of Washington, Physics-Astronomy Building, Box 351580, Seattle, WA 98195-1700, USA}
\affiliation{eScience Institute, University of Washington, Physics-Astronomy Building, Box 351580, Seattle, WA 98195-1700, USA}

\author[0000-0003-1370-5010]{Michael D. Gladders}
\affiliation{Department of Astronomy and Astrophysics, University of Chicago, 5640 South Ellis Avenue, Chicago, IL 60637, USA}
\affiliation{Kavli Institute for Cosmological Physics, University of Chicago, 5640 South Ellis Avenue, Chicago, IL 60637, USA}

\author[0000-0002-7559-0864]{Keren Sharon}
\affiliation{Department of Astronomy, University of Michigan, 1085 S. University Ave, Ann Arbor, MI 48109, USA}

\author[0000-0002-7113-0262]{Viraj Manwadkar}
\affiliation{Department of Physics, Stanford University, Stanford, CA 94305}
\author[0000-0001-6454-1699]{Yunchong Zhang}
\affiliation{Department of Physics and Astronomy and PITT PACC, University of Pittsburgh, Pittsburgh, PA 15260, USA}

\author[0000-0002-0243-6575]{Jacqueline Antwi-Danso}
\affiliation{Dunlap Institute for Astronomy and Astrophysics, 50 St. George Street, Toronto, Ontario, M5S 3H4, Canada}
\affiliation{Department of Astronomy, University of Massachusetts, Amherst, MA 01003, USA}
\affiliation{David A. Dunlap Department of Astronomy and Astrophysics, University of Toronto, 50 St. George Street, Toronto, Ontario, M5S
3H4, Canada}

\author[0009-0000-5333-9970]{Dylan Berry}
\affiliation{Department of Astronomy \& the DiRAC Institute, University of Washington, Physics-Astronomy Building, Box 351580, Seattle, WA 98195-1700, USA}

\author[0000-0001-5063-8254]{Rachel Bezanson}
\affiliation{Department of Physics and Astronomy and PITT PACC, University of Pittsburgh, Pittsburgh, PA 15260, USA}

\author[0000-0003-2200-5606]{H{\aa}kon Dahle}
\affiliation{Institute of Theoretical Astrophysics, University of Oslo, P.O. Box 1029, Blindern, NO-0315 Oslo, Norway}
%\email{cleri@psu.edu}

\author[0009-0004-1123-001X]{Carlos Garcia Diaz}
\affiliation{Department of Astronomy, University of Massachusetts, Amherst, MA 01003, USA}

\author[0000-0001-5341-2162]{Darko Donevski}
\affiliation{National Center for Nuclear Research, Pasteura 7, 02-093 Warsaw, Poland}
\affiliation{SISSA, Via Bonomea 265, 34136 Trieste, Italy}
\affiliation{INAF, Osservatorio Astronomico di Trieste, via Tiepolo 11, I-34131, Trieste, Italy}

\author[0009-0009-9810-3334]{Alex~J.~Geiger}
\affiliation{Department of Physics and Astronomy and George P. and Cynthia Woods Mitchell Institute for Fundamental Physics and Astronomy, Texas A\&M University, 4242 TAMU, College Station, TX 77843-4242, US}

\author[0000-0001-6763-5551]{Massissilia L. Hamadouche}
\affiliation{Department of Astronomy, University of Massachusetts, Amherst, MA 01003, USA}

\author[0000-0001-6755-1315]{Joel Leja}
\affiliation{Department of Astronomy and Astrophysics, The Pennsylvania State University, University Park, PA 16802, USA}
\affiliation{Institute for Computational \& Data Sciences, The Pennsylvania State University, University Park, PA 16802, USA}

\author[0000-0003-3266-2001]{Guillaume Mahler}
\affiliation{STAR Institute, Quartier Agora - Allée du six Août, 19c B-4000 Liège, Belgium}

\author[0000-0001-8592-2706]{Alexandra Pope}
\affiliation{Department of Astronomy, University of Massachusetts, Amherst, MA 01003, USA}

\author[0000-0003-3256-5615]{Justin Spilker}
\affiliation{Department of Physics and Astronomy and George P. and Cynthia Woods Mitchell Institute for Fundamental Physics and Astronomy, Texas A\&M University, 4242 TAMU, College Station, TX 77843-4242, US}

\author[0000-0002-1106-4881]{Ezra Sukay}
\affiliation{Department of Physics and Astronomy, Johns Hopkins University, 3400 N. Charles Street, Baltimore, MD 21218, USA}

\author[0000-0001-6477-4011]{Francesco Valentino}
\affiliation{Cosmic Dawn Center (DAWN), Denmark}
\affiliation{ DTU Space, Technical University of Denmark, Elektrovej 327, DK-2800 Kgs. Lyngby, Denmark}

\begin{abstract}
The interstellar medium of massive quiescent galaxies holds key clues to how they quench, yet remain poorly understood. Although dust is expected to be depleted rapidly ($\lesssim$100 Myr) after quenching, some quiescent galaxies host substantial dust reservoirs, raising the question of whether dust survives quenching or is replenished afterward. We present ALMA Band 6 continuum observations ($\lambda_{\mathrm{rest}}$$>$250$\mu$m) of eight strongly-lensed, spectroscopically-selected quiescent galaxies at z$\sim$1-2.5 from the COOL-LAMPS and REQUIEM surveys, plus five archival REQUIEM systems. The sample spans more than two orders of magnitude in dust-to-stellar-mass ratio ($f_{\mathrm{dust}}<10^{-5}$ to $>10^{-3}$), and two galaxies host extended, resolved dust. Comparing $f_{\mathrm{dust}}$ with time since quenching, roughly half the sample, preferentially at higher redshift, is consistent with rapid quenching and harsh dust destruction, while lower-redshift, more slowly quenched systems are consistent with milder destruction, possibly reflecting an evolving dominant quenching mode. The two most dust-rich galaxies exceed even the mildest model destruction tracks, with J1201-3952 retaining the highest $f_{\mathrm{dust}}$ $\sim$1.8 Gyr after quenching. If these systems are truly quiescent, their dust masses approach maximum stellar yields, pointing to ISM grain regrowth and/or TP-AGB dust production. However due to our single-band continuum data, $f_{\mathrm{dust}}$ is uncertain ($\pm$0.3 dex systematic uncertainty from the unknown dust temperature) and we cannot rule out obscured star formation in these two galaxies. Our results suggest that dust survival depends on how galaxies quench: rapid, harsh quenching efficiently destroys dust, while slower, milder quenching can leave dust reservoirs that persist or are replenished long after star formation ceases.
\end{abstract}

\section{Introduction} \label{sec:intro}

In the current picture of massive galaxy evolution, a large fraction of massive galaxies exhibit low star formation rates \citep[SFRs; e.g.,][]{muzzin2013a,ilbert2013,Tomczak2014,Davidzon2017}, compact, bulge-dominated morphologies \citep[e.g.,][]{daddi2005, vandokkum2008,vanderwel2014}, and red colors from old stellar populations \citep[e.g.,][]{kriek2006,williams2009,whitaker2011,whitaker2013, muzzin2013}. These quiescent galaxies are thought to evolve from star-forming disk galaxies, but the physical processes through which star formation shuts down and the origin of their compact morphologies is unclear \citep[for recent comprehensive reviews, see][]{whitaker2026,bluck2026a, bluck2026b}. Proposed quenching mechanisms must either heat, remove, consume, or expel reservoirs of cold molecular gas, which serve as the fuel for star formation, either over rapid ($<$1Gyr) or slower ($>$1Gyr) timescales \citep[e.g.,][]{Carnall2018,Park2024, edward2026}. Therefore, the joint analysis of star formation histories (SFH) and cold dust and gas content in quiescent galaxies can reveal how, when, and where quenching happens. 

Several studies have sought to study molecular gas in quiescent and post-starburst galaxies, primarily utilizing CO spectroscopy \citep{hayashi2018,spilker2018,woodrum2022,wu2023,zanella2023,donofrio2025,setton2025, scholtz2026}. However, a challenge of this approach is the long integration times required to detect CO, particularly in systems that are expected to be intrinsically gas-poor if quenching has occurred through gas depletion. Consequently, CO has been detected in only a handful %(<XX number) 
of quiescent galaxies, limiting robust constraints on the molecular gas content of the population. Beyond total gas content, spatially-resolved gas measurements can identify specific quenching mechanisms and elucidate how stellar mass assembles across different regions of a galaxy. For example, centrally depleted gas reservoirs could indicate inside-out quenching driven by AGN feedback or morphological quenching, consistent with the suppressed central star formation and older central stellar populations seen in resolved studies of massive galaxies at cosmic noon \citep{tacchella2015, nelson2016,tacchella2018, akhshik2023}. Similarly, gas displaced from the stellar body or swept up in outflows would point to environmental stripping or ejective feedback \citep{spilker2022, belli2024, deugenio2024}. However, such resolved CO observations are prohibitively expensive for most quiescent galaxies at $z>1$.

A less expensive tracer of cold gas is dust continuum at $\lambda_{\mathrm{rest}}>250\mu$m, which is closely correlated with CO luminosity in star-forming galaxies \citep{scoville2014,scoville2016}. While dust continuum has been used to estimate molecular gas mass in local quiescent galaxies \citep{gomez2010,rowlands2012,Lianou2016,michalowski2019,lesniewska2023,michalowski2024}, studies at higher redshift often result in non-detections \citep{caliendo2021,kalita2021,whitaker2021, suzuki2022, chworowsky2024,umehata2025,siegel2025,spilker2025,chang2026}, indicating that little cold dust is present. There are several individual detections of dust continuum in quiescent galaxies; some of these systems are strongly gravitationally lensed, enabling observations of less massive dust reservoirs \citep{whitaker2021,morishita2022}, or they probe the dustiest subset of the quiescent population \citep{donevski2023, lee2024}. Alternatively, stacking has been used to measure dust masses below the detection limit for statistical samples of quiescent galaxies. Early studies showed that stacking recovered  higher dust masses than individual measurements \citep{gobat2018, magdis2021,blanquezsese2023}, whereas recent mass-complete ALMA stacks found lower average $f_{\mathrm{dust}}$ values consistent with individual measurements \citep{Adscheid2025}.

This tension between stacking and individual measurements of dust mass may partly reflect methodological biases. Stacks can be boosted by the dustiest sources \citep{lorenzon2025}, by source confusion in low-resolution far-IR maps \citep{Adscheid2025}, or by including lower-mass galaxies with higher $f_{\mathrm{dust}}$ fractions \citep{blanquezsese2023}, while individual detections remain limited to a handful of objects.  Dust mass estimates from both approaches are further uncertain due to the assumed dust temperature \citep{cochrane2022}. However, the wide range of measured dust masses may also reflect genuine diversity in the ISM of quiescent galaxies. Supporting this point, simulations predict gas-to-dust ratios spanning up to four orders of magnitude \citep{whitaker2021b,lorenzon2025}, and observations find unusually high and varied gas-to-dust ratios in some systems \citep{spilker2025,lorenzon2025b}.  

Dust is injected by supernovae and the winds of evolved stars and is subsequently reprocessed in the ISM, where grains grow in dense gas and are shattered into smaller grains in diffuse gas \citep{salim2020, narayanan2026, parente2026}. Once star formation ceases, these sources fade, with Type II supernovae ending within $\sim$40 Myr, dust production by asymptotic giant branch (AGB) stars declining over the following several hundred Myr to $\sim$2 Gyr as progressively lower-mass stars reach this phase, and grain growth stalling as cold, dense gas is consumed, heated, or expelled. Destruction, by contrast, can be rapid, through thermal sputtering in hot gas, supernova shocks, astration, and outflows \citep{li2019}, with sputtering alone destroying grains within $\sim$1-10 Myr in hot halos \citep{dwek1980, jones1994, jones1996, hirashita2017}. Quiescent galaxies are therefore naively expected to lose their dust shortly after quenching. Yet observations of quiescent galaxies with substantial dust reservoirs \citep{hayashi2018, gobat2018, blanquezsese2023, lee2024} suggest that removal may be prolonged in some quiescent systems, or offset by replenishment through grain growth or asymptotic giant branch (AGB) production  \citep{donevski2023,lorenzon2025, donevski2026}.

Beyond the timescale of dust destruction, a second confounding factor is the diversity of SFHs among quiescent galaxies, which encode a wide range of quenching timescales and times since quenching. While only $<10$\% of quiescent galaxies at $z\sim1$ are predicted to be recently quenched, this number rises to $\sim50$\% at $z\sim2$ \citep[e.g.,][and references therein]{whitaker2026}. If dust destruction proceeds more slowly than quenching, recently quenched galaxies may retain dust formed before quenching, or sustain enough residual star formation to continue producing it. Thermally pulsing asymptotic giant branch (TP-AGB) stars add a further channel: they are expected to be prominent in quiescent galaxies with intermediate-age stellar populations ($\sim$0.3-2 Gyr), particularly at $z\sim1$, and could produce dust long after quenching, which would trace the intermediate-age stars rather than the dense core \citep{lu2025}. 

The details of dust destruction after star formation has been disrupted remain poorly understood.  \citet{donevski2026} predict that the fate of the dust depends most strongly on the local ISM conditions, ranging from harsh conditions, with or without AGN feedback, that rapidly destroy dust, to milder conditions that prolong its survival. In some cases, particularly in low- to intermediate-mass quiescent galaxies, grain regrowth keeps the dust mass roughly constant \citep{donevski2026, lorenzon2025}. This heterogeneity may make dust continuum an unreliable tracer of molecular gas in quiescent galaxies \citep{whitaker2021b, spilker2025}.  However, because dust content is tied to both quenching conditions and stellar population age, the dust continuum is itself a powerful probe of quenching, complementary to molecular gas studies.

In this paper, we present new ALMA Band 6 continuum observations of a sample of eight strongly lensed,  spectroscopically-selected quiescent galaxies from the COOL-LAMPS \citep{Khullar2021} and REQUIEM surveys, and analyze their cold dust properties alongside SFHs derived from spectro-photometric spectral energy distribution (SED) fitting. We combine this sample with archival ALMA observations of five REQUIEM galaxies \citep{whitaker2021,akhshik2023}. Lensing magnification boosts both sensitivity, making it possible to detect, or place stringent limits on, the faint dust reservoirs of individual quiescent galaxies at $z>1$ \citep[e.g.,][]{whitaker2021} that are otherwise only accessible through stacking.  Where dust is detected, lensing also stretches the emission across the sky, so even our low-resolution ($\sim$1$''$) data can begin to distinguish extended from compact dust, and this sets up higher-resolution follow-up to map the dust relative to the stars.

The paper is structured as follows. Section \ref{sec:Data} describes the sample, existing optical data, and the reduction of the ALMA dust continuum and ground-based spectroscopy. Section \ref{sec:methods} presents our method for determining lensing magnification, measuring dust masses, fitting SEDs, and deriving quenching timescales from SFHs. Section \ref{sec:results} relates the dust properties of the sample to redshift, specific star formation rate (sSFR), stellar age, and SFH. In Section \ref{sec:discussion}, we examine the relationship between time since quenching and dust fraction, what it implies about changing modes of dust destruction across the redshift range of the sample, and whether the largest dust masses arise from stellar dust production (AGB stars and Type II supernovae) or grain growth in the ISM, as well as whether the dust-rich systems could host obscured star formation. We summarize our key findings in Section~\ref{sec:conclusion}.

\section{Data} \label{sec:intro}

\label{sec:Data}

Our sample consists of 13 strongly lensed quiescent galaxies at
1$<$$z$$<$3, including seven from the ChicagO Optically-selected strong Lenses -- Located At the Margins of Public Surveys (COOL-LAMPS) collaboration \citep{Khullar2021} and six from the REsolving QUIEscent Magnified (REQUIEM) Galaxies Survey \citep{whitaker2021, akhshik2023}. We present new dust continuum imaging for 8 of the 13 galaxies (Section~\ref{sec:alma}, see Figure \ref{fig:rgb}); 7 of these galaxies are from COOL-LAMPS. The COOL-LAMPS and REQUIEM samples differ in their imaging depth and resolution (Sections~\ref{sec:coollamps} and \ref{sec:requiem}) and in their rest-frame wavelength coverage (Section~\ref{sec:sed_comparison}). While different tools are used to model their stellar populations and derive star formation histories (SFHs), the underlying assumptions are
similar (Section~\ref{sec:sed_comparison}). We describe each below and summarize the differences in Section~\ref{sec:sed_comparison} and Table~\ref{tab:sample}.

\subsection{COOL-LAMPS}\label{sec:coollamps}
\subsubsection{Sample Selection}

COOL-LAMPS candidates were identified through visual inspection of image cutouts from the DESI Legacy Imaging Surveys DR9 (DECaLS;
\citealp{Dey_2019}) and DELVE \citep{alexdw_delve_2021}, centered on red-sequence galaxies and galaxy clusters. Members of the collaboration ranked candidates by the likelihood of being a strong lens, prioritizing systems where background arcs lie at the margins of color-color and color-magnitude spaces (i.e., bright and red objects in $grz$ co-adds-based photometric catalogs), and the highest-ranked systems were targeted for spectroscopic follow-up with Magellan (Section~\ref{sec:coollamps_spectra}). From
this parent sample we include the seven systems for which follow-up ALMA dust continuum imaging was acquired and whose lensed source shows a spectrum consistent with a quiescent stellar population (e.g., strong Balmer/4000 $\mathrm{\AA}$ break and absorption features, with weak or absent
emission lines). A detailed study of one of these systems, COOL\,J1323+0343, is presented in \citet{sukay2022}, with the full sample (data and stellar population synthesis modeling) to be presented in Khullar et al. (in prep). Figure~\ref{fig:rgb} shows RGB composite images of these seven sources (in addition to J1522-2535, to be presented in Section~\ref{sec:requiem}).

\subsubsection{Optical and Near-Infrared Imaging and Modeling} \label{sec:coollamps_images}

Optical $grz$ imaging was obtained from DECaLS DR9 \citep{Dey_2019}.  
For a subset of targets, we obtained follow-up $JHKs$ imaging with the FourStar Infrared Camera (FOURSTAR; \citealp{Persson_2008}) on the Magellan/Baade telescope. These data were reduced to astrometrically and photometrically calibrated stacks using a custom pipeline built via IRAF (\citealp{Tody1986}, \citealp{Tody1993}) and PHOTPIPE (\citealp{Rest_2005}, \citealp{Garg_2007}, \citealp{Miknaitis_2007}). The imaging reduction and photometric analysis broadly
follow \citet{sukay2022}, with COOLJ1323+0343 serving as a representative example of the pipeline. The full details for the sample will appear in Khullar et al. (in prep). Typical seeing was 0.7$''$ in the optical and 1$''$ in the near-IR, and all targets were resolved due to their strongly lensed nature. 

There is significant contamination in the arcs due to flux from foreground/interloper cluster member galaxies that form the gravitational lens. Therefore, we measure photometry via \texttt{GALFIT} modeling \citep{Peng_2002}. We fit the lens galaxy and nearby contaminants with S{\'e}rsic profiles and model the arc as multiple S\'ersic components in across each available image. The model components belonging to the primary arc are merged into a single
nominal ``banana'' aperture whose outer boundary traces the modeled arc \citep[see also,][]{Khullar2021,sukay2022,akhshik2023}. The same aperture is effectively applied to all bands, and fluxes are measured by adding the inferred flux from all \texttt{GALFIT} model components 9in lens-subtracted images. The background is estimated by placing multiple apertures in blank sky with the same ``banana'' aperture as the science target.

\subsubsection{Ground-Based Spectroscopy}\label{sec:coollamps_spectra}

Spectra of the lensed sources were obtained with the Low Dispersion Survey Spectrograph 3 (LDSS3) on the Magellan/Clay telescope between 2021-2024, using the 1$''$ x 4$'$ longslits and the VPH-Red Grism disperser, with total integration times of 1--2\,hrs (except COOL J1323+0343, where the spectrum was observed with NOT/ALFOSC; \citealt{sukay2022}). 
Slits were aligned along the brightest arc segment where possible. We reduced the data with a custom IRAF pipeline that performed bias subtraction, flat-fielding, wavelength calibration, sky subtraction, and optimal extraction of the 1D spectrum; we used imaging for relative and absolute flux calibration as a parameter in spectral energy distribution fitting (Section \ref{sec:coollamps_images}). Spectroscopic redshifts were measured via \texttt{Prospector}, with typical uncertainties of $\delta z \sim$ 0.05.  The wavelength coverage of rest-frame 3500-4300 $\mathrm{\AA}$ provides coverage of the [OII] emission line, and Balmer absorption features like H$\gamma$,   H$\delta$,  H$\epsilon$,  H7, H8, and the Ca K\&H doublet absorption features.

\subsection{REQUIEM}\label{sec:requiem}

The REsolving QUIEscent Magnified (REQUIEM) galaxies survey targets eight strongly lensed quiescent galaxies at $1.6<z<2.9$ with demagnified stellar masses of $10.6<\log(M_\star/M_\odot)<11.2$ \citep{akhshik2023,Newman2026}\footnote{Updates to the lens model of MRG-M0138 with JWST-era data and new constraints lower the estimated stellar mass to 11.34 (12.82 without accounting for lensing) \citep{Newman2026}.}. We include 7 of these galaxies in our sample, including new dust continuum imaging for J1522-2535 (Figure~\ref{fig:rgb}; see Section~\ref{sec:alma}); we exclude SDSSJ0851 \citep{akhshik2020} as the ALMA dust continuum observations were not executed. The REQUIEM-2D data consist of deep (5--15 orbit) HST/WFC3 G141 grism spectroscopy (HST-GO-15663) together with archival HST WFC3 and ACS imaging spanning F336W--F160W and Spitzer/IRAC 3.6 and 4.5 $\mu$m photometry, all reduced with \texttt{Grizli} \citep{grizli2019}. Contaminating neighbors are modeled with \texttt{GALFIT} in F160W and subtracted, and fluxes are measured in ``banana''-shaped object apertures that follow the arc's axis of symmetry \citep{Wuyts2010, akhshik2023}, an approach analogous to the \texttt{GALFIT}-based arc apertures used for COOL-LAMPS (Section~\ref{sec:coollamps_images}).

\subsection{ALMA Dust Continuum Imaging} \label{sec:alma}

\begin{figure*}
    \centering
    \includegraphics[width=\textwidth]{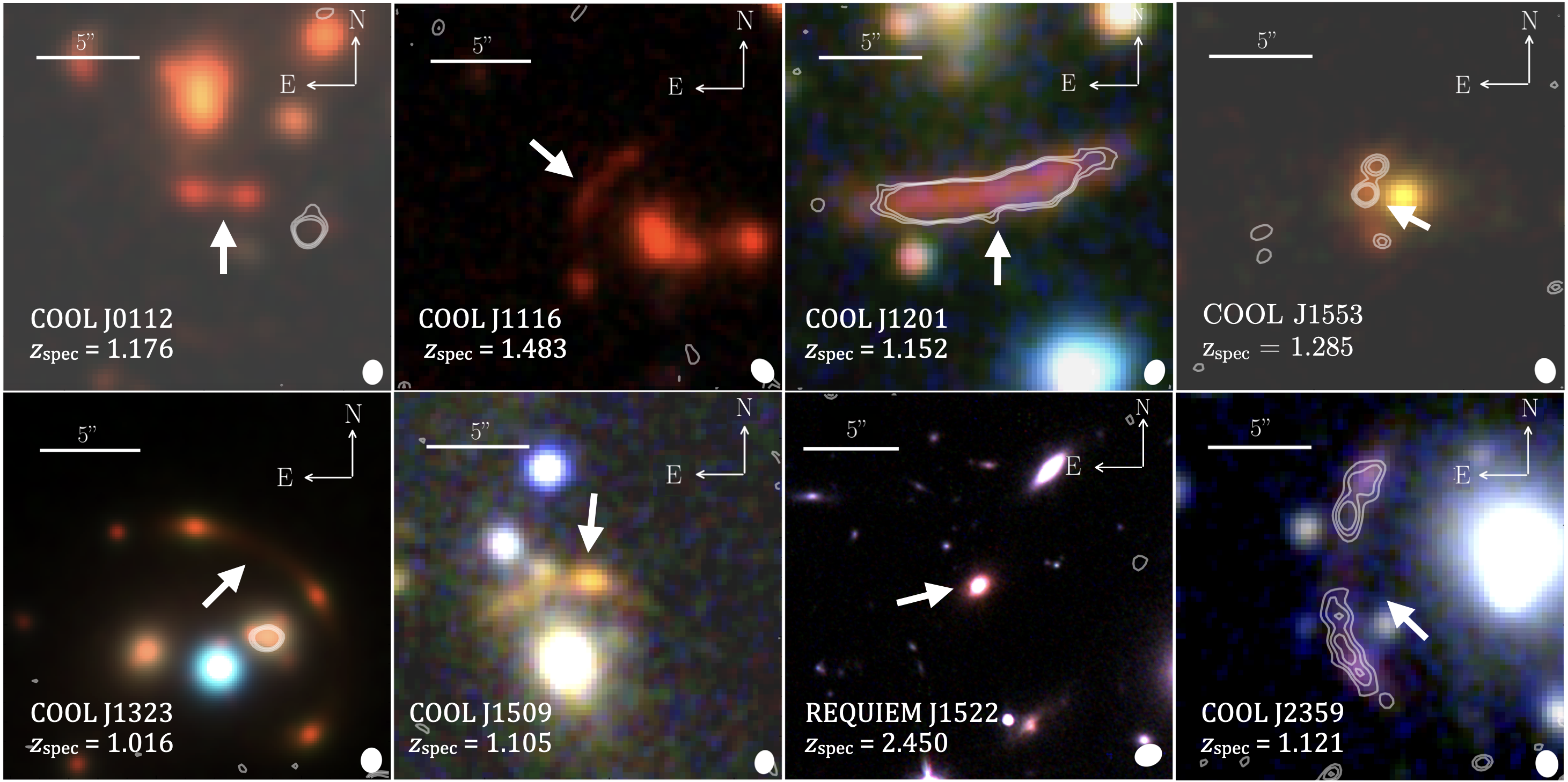}
    \caption{RGB images for the sample of eight lensed, spectroscopically-selected quiescent galaxies from the COOL-LAMPS and REQUIEM surveys. The optical imaging is $griz$ with the exceptions of (i) J1201-3952, where we use H-band data, (iii) J1323-0343, where we use F140W data \citep{Zhuang_AGEL_2023}, and (iii) J1522-2525, for which we use HST imaging (F110W, F125W, and F140W). We show 3, 4, and 5$\sigma$ contours from the 1.3mm dust continuum. Arrows point to the lensed arcs; in the case of J1323-0343, there is a nearby 5$\sigma$ source which is associated with the arc.} 
    \label{fig:rgb}
\end{figure*}

The ALMA continuum observations used in this study were taken as part of program 2023.1.00430.S and combined with the archival programs 2018.1.00276.S and 2019.1.0027.S presented in \citet{whitaker2021}. The dust continuum observations in this program are centered at 226GHz in Band 6, corresponding to an observed wavelength of 1.3mm ($\sim$600$\mu$m at $z=1.2$) covering the 7.5GHz bandwidth in the Time Division Mode (TDM). In order to make a global measurement of dust mass, lower resolution data is optimal. Given the compact nature of dust in quiescent galaxies from previous studies \citep[e.g.][]{whitaker2021},  the antenna configuration was chosen to obtain low spatial resolution dust continuum ranging from 1\arcsec-1.5\arcsec. The baselines range from 15.1 m to 313.7 m with 43-45 antennae. On source time was 6.8 hours.

The ALMA data are reduced and calibrated using CASA \citep{casa} version 6.5.4.9. We use pipeline-calibrated visibility data for the continuum imaging, generated with the CASA task \texttt{tclean}. The pipeline cleaning routine is performed down to 2$\sigma$ noise level with Briggs weighting and a robust parameter set to $R=0.5$. With this weighting scheme, the median circularized beam in the dataset has a radius of 1.12\arcsec (ranging from 1.04\arcsec to 1.28\arcsec) and the typical RMS noise is 20 $\mu$Jy/beam. For targets that are undetected in the pipeline imaging, we test whether cleaning in interactive mode with natural weighting ($R=2.0$) and $u-v$-tapering to 2\arcsec (a physical scale of 16.5 kpc at $z\sim1$, or $\sim3$kpc assuming $\mu$=10) recovers signal in the case where the dust morphology is extended. We find no extended emission in these sources and therefore default to the pipeline imaging scheme as it offers better sensitivity.

Out of the eight targets observed in program 2023.1.00430.S (Figure \ref{fig:rgb}), we identify three targets with S/N $>$ 5 dust continuum detection: J1201-3952, J1553-0142, J2359-0208. We fit J1553-0142 with a Gaussian using CARTA and find that the FWHM is smaller than the beam size; therefore we conclude that this source is unresolved, whereas the dust emission in J1201-3952 and J2359-0208 is extended. J1201-3952 has a sufficiently high S/N ($>$15) 
to be flagged by CASA as a self-calibration candidate.

\begin{table*}[t]
    \centering
    \fontsize{8}{9}
    \selectfont
    \begin{tabular}
    {|cccccccc|}
        \hline
       ID  & \emph{z}$_\mathrm{spec}$ & $\mu_{50}$ & S$_{1.3\mathrm{mm}}$ [mJy] & log(M$_\mathrm{\mathrm{star}}$/M$_\odot$) & log(M$_\mathrm{\mathrm{dust}}$/M$_\odot$) & Age [Gyr] & sSFR [yr$^{-1}$]
       \\ \hhline{|========|}
       J1323-0343 & 1.016 & 15 $\pm$ 1 & $<$0.04 & 11.28$_{-0.09}^{+0.1}$ & $<$6.5 & 3.8$_{-0.9}^{+0.6}$&-13.1$_{-1.4}^{+0.7}$\\
       J2359-0208 & 1.121 & 15 $\pm$ 5 & 1.36$\pm$0.07 & 11.00$_{-0.2}^{+0.2}$ & 8.1$\pm$0.1 &3.8$_{-0.4}^{+0.3}$ &-10.67$_{-0.09}^{+0.06}$\\
       J1201-1443 & 1.152 & 15 $\pm$ 5 & 4.07$\pm$0.16 & 11.2$_{-0.1}^{+0.2}$ & 8.6$\pm$0.1 & 4.6$_{-0.3}^{+0.2}$ & -10.68$_{-0.06}^{+0.7}$\\
       J0112-2053 & 1.176 & 20 $\pm$ 10 & $<$0.05 & 11.0$_{-0.3}^{+0.3}$ & $<$6.5 & 4.92$_{-0.01}^{+0.05}$ &-12.3$_{-0.6}^{+0.3}$\\
       J1509-3308 & 1.105 & 20 $\pm$ 10 & $<$0.04 & 10.9$_{-0.2}^{+0.2}$ & $<$6.5 & 4.7$_{-0.4}^{+0.2}$ &-11.8$_{-0.1}^{+0.1}$\\
       J1553-0142 & 1.285 & 3.5 $\pm$ 1.5 & 0.24$\pm$0.05 & 12.1$_{-0.2}^{+0.2}$ & 8.0$\pm$0.2 & 4.39$_{-0.3}^{+0.06}$&-11.6$_{-0.7}^{+0.3}$\\
       J1116-1443 & 1.483 & 10 $\pm$ 3 & $<$0.04 & 11.24$_{-0.1}^{+0.1}$ & $<$6.7 & 3.8$_{-0.2}^{+0.2}$&-11.1$_{-0.5}^{+0.5}$\\
       J1522-2535$^a$ & 2.450 & 15.4 $\pm 6.2 $ & $<$0.03 & 10.6$\pm0.2$ & $<$6.4 & 0.61$\pm0.06$ & $<$-11.77
       \\ \hline
    \end{tabular}
    \caption{Sample summary for strong lens quiescent galaxies where we add new ALMA dust continuum imaging (Section~\ref{sec:alma}). $^a$J1522-2535 is from the REQUIEM survey; stellar mass, sSFR, and age for J1522-2535 come from \citet{newman2018}.}
    \label{tab:sample}
\end{table*}

Following Equation 7.6 of the ALMA Cycle 12 Technical Handbook, the maximum recoverable scale ranges from 10\arcsec  to 10.8\arcsec. Dust emission more extended than this maximum recoverable scale is not recovered by these observations; since the major axes of J1201-3952 and J2359-0208 are 8\arcsec and 13\arcsec, we expect that we should recover most of the flux density from these sources. However, for J1201-3952 and J2359-0208, we test cleaning in interactive mode with natural weighting ($R=2.0$) and $u-v$-tapering to 2\arcsec (a physical scale of 16.5 kpc at $z\sim1$, or $\sim3$kpc assuming $\mu$=10) and 3\arcsec (a physical scale of 25 kpc at $z\sim1$, or $\sim5$kpc assuming $\mu$=10) to determine whether we are resolving out any flux. In the case of J2359-0208, we find that the flux density after using natural weighting increases by $<1$\% and natural weighting with $u$-$v$-tapering to 2\arcsec decreases the flux density 17\%; therefore, we conclude that natural weighting is sufficient to recover extended emission. For J1201-3952, natural weighting and $u$-$v$-tapering to 2\arcsec increases the flux density $\sim$15\%.

\section{Methods and Analysis}

\label{sec:methods}

\subsection{Gravitational Lensing Magnification}
\label{sec:lensmodel}

As with the optical photometry and spectroscopy, the lens modeling of the COOL-LAMPS systems from ground-based imaging is described in full in Khullar et al. (in prep). We briefly summarize the procedure here. Lens models are constructed with the parametric lens modeling algorithm \texttt{Lenstool} \citep{Jullo_2007}, which identifies the best-fit and samples the posterior via Markov Chain Monte Carlo (MCMC). The mass distribution of each lens is parameterized as dual pseudo-isothermal ellipsoid halos (dPIE, \citealp{Eliasdottir_2007}), constrained by multiple-image systems identified in the $JHKs$ imaging where available (otherwise in $grz$ imaging).

Using the apertures defined from the optical imaging (Section~\ref{sec:coollamps}), we determine the number and spatial extent of the lensed images that make up each primary arc, and use these to compute an effective magnification, $\mu$, for the region of the source contributing to our photometry and spectroscopy. We define $\mu$ as the harmonic mean of the magnification over the pixels within the aperture (as is standard practice; e.g. see \citealt{sharon2020}). By drawing 1000 posterior chains from the \texttt{Lenstool} output, we report the 16th, 50th, and 84th percentiles of the resulting $\mu$ distribution in Table~\ref{tab:sample}; we note here that ground imaging-based lens models are severely limited in the ability to constrain multiple images at various redshifts and spectral energy distributions (SEDs), and hence the magnification uncertainties are likely biased and/or the estimated median value is biased.

Strong lensing is wavelength-independent, so quantities derived from spectral shape such as dust attenuation and stellar age are roughly independent of magnification, as is the specific star formation rate, in which the magnification cancels out. Absolute quantities such as stellar mass and star formation rate are scaled by $\mu$. To propagate the uncertainties in both the stellar population fits and the lens models, we construct the delensed distributions by Monte Carlo sampling. First we make 1000 draws from the stellar mass and star formation rate posteriors, and then, independently, we draw from the magnification distribution. Each of these physical quantities is divided by the drawn $\mu$ (considering the magnification boost to be a multiplicative factor). The resulting delensed stellar masses and star formation rates are reported in Table~\ref{tab:sample}. We note that the stellar mass of J1553-0142 may be overestimated due to uncertainty in the lensing model.

For the REQUIEM targets, we adopt the magnifications compiled in Table 4 of
\citet{akhshik2023} and the references therein (estimated using the same process as COOL-LAMPS), with one exception. For MRG-M0138, we replace the pre-JWST value of $\mu = 13 \pm 5$ \citep{newman2018, rodney2021} with $\mu = \mathrm{29}$, derived from recent lens models that incorporate new JWST imaging and VLT/MUSE spectroscopic constraints \citep{acebron2025, ertl2025, suyu2026}.

\subsection{Modeling Stellar Populations}
\label{sec:sed}

\subsubsection{COOL-LAMPS}
\label{sec:sed_coollamps}
We derive physical properties (e.g., stellar mass, SFR, age) and non-parametric SFHs for the COOL-LAMPS galaxies by jointly fitting the ground-based spectroscopy and the multi-band photometry ($grizJHK_s$, or the subset available for each target) measured within the apertures described in Section~\ref{sec:coollamps_images}, using the Bayesian stellar population synthesis (SPS) framework \texttt{prospector} \citep{leja2017, johnson2021}. Stellar populations are modeled with FSPS \citep{conroy2009, conroy2010} using MIST isochrones \citep{choi2016} and the MILES spectral library \citep{sanchezblazquez2006}, assuming a \citet{chabrier2003} initial mass function. Nebular line and continuum emission are included using the CLOUDY-based models embedded in FSPS \citep{byler2017, byler}.

We adopt a non-parametric SFH with 10 fixed lookback-time bins and the continuity prior of \citet{leja2019}, which is placed on the ratio of star formation rates in adjacent bins and favors smooth transitions between them. The first two bins span 0--30,Myr and 30--100,Myr, the last bin spans $0.9$--$1.0\,t_{\rm univ}$, where $t_{\rm univ}$ is the age of the Universe at the observed redshift, and the remaining seven bins are spaced evenly in logarithmic time between 100,Myr and $0.9\,t_{\rm univ}$. Dust attenuation follows the \citet{kriek2013} law, with the diffuse optical depth at 5500 $\mathrm{\AA}$ (${\tau}_2$, or equivalently $A_V$) and the power-law slope offset ($n$, \texttt{dust\_index}) as free parameters; the strength of the 2175 $\mathrm{\AA}$ bump is tied to $n$ \citep{noll2009, kriek2013}. We adopt uniform priors of ${\tau}_2 \in [0, 4]$ and $n \in [-1.0, 0.4]$ \citep{leja2017, leja2019}. Stellar metallicity is a free parameter, with flat priors in log-space (based on the limits of the MILES spectral library within FSPS). 

In our joint spectro-photometric fit, we utilise a fifth-order Chebyshev polynomial for relative and absolute flux-calibration vector, marginalizing over systematic wavelength-dependent differences between the spectroscopic and photometric fluxes. We also fit for the redshift within $[z_{\rm spec} - 0.05,\, z_{\rm spec} + 0.05]$ to allow for uncertainties in the wavelength calibration of the ground-based spectra. The posterior is sampled with \texttt{dynesty} \citep{speagle2020}. ALMA observations are not included in the fits, so no energy-balance assumptions enter our fiducial SPS model.

Because magnification enters only as a normalization of the spectrophotometric data, the inferred SFH shape is independent of the lens model, while extensive quantities are corrected using the magnifications from Section~\ref{sec:lensmodel} (post-fitting). We primarily rely on ratios (e.g., specific SFR, dust-to-stellar mass ratio), for which the uncertain effect of magnification is minimized. Further details of the SFH modeling and its caveats are given in \citet{sukay2022} and Khullar et al. (in prep).

\subsubsection{REQUIEM}
\label{sec:sed_requiem}

For the REQUIEM galaxies we adopt the SFHs of \citet{akhshik2023}, derived with the \texttt{requiem2d} code \citep{akhshik2020}, which jointly fits the multi-band HST photometry and the 2D G141 grism spectra in their native 2D space, with a multiplicative polynomial fit alongside the stellar population model. While \citet{akhshik2023} focus on spatially resolved stellar populations, we use only their global SFHs (summed over all spatial bins), which additionally incorporate the unresolved Spitzer/IRAC constraints, to match the integrated COOL-LAMPS measurements. As for COOL-LAMPS, the models use FSPS templates with a \citet{chabrier2003} IMF and a non-parametric SFH with a continuity prior, here with 20 lookback-time bins. Dust attenuation follows the same \citet{noll2009}/\citet{kriek2013} parameterization, but with the birth-cloud (${\tau}_1$), diffuse (${\tau}_2$), and slope ($n$) parameters all free. Metallicity and ${\tau}_2$ priors are set by first fitting the photometry with \texttt{Prospector}-$\alpha$ \citep{leja2017}, dividing the resulting posteriors into 16 template boxes, and refitting the box weights simultaneously with the SFH, such that the \texttt{Prospector} posteriors act as weakly informative priors. The posterior is sampled with the No-U-Turn Sampler \citep{hoffman2014} in \texttt{pymc3} \citep{salvatier2016}.

\subsubsection{Comparison of Modeling Assumptions}
\label{sec:sed_comparison}
Both subsamples are modeled with FSPS stellar populations, a \citet{chabrier2003} IMF, non-parametric SFHs with a continuity prior, and a \citet{kriek2013} dust law. The main differences are the SFH time resolution (10 bins for COOL-LAMPS versus 20 for REQUIEM); the treatment of young-star attenuation (fixed at twice the diffuse attenuation versus a free birth-cloud component); the metallicity prior (a \citet{gallazzi2005} mass--metallicity prior versus a weakly informative prior derived from photometry-only fits); the sampler (\texttt{dynesty} versus NUTS); and the spectroscopic constraint (ground-based spectra versus HST/G141 grism spectra). Since non-parametric SFH constraints are prior-sensitive, particularly in the oldest bins \citep{leja2019}, we compare the subsamples using derived quantities that are less sensitive to bin choice, namely the quenching timescale and the time elapsed since quenching, defined following \citet{whitaker2026}. We note that \citet{akhshik2023} report
stellar masses 0.1--0.5\,dex higher than earlier parametric fits to the same galaxies \citep{newman2018, man2021}, an offset attributed both to the non-parametric SFHs and to their aperture photometry; the COOL-LAMPS masses should be similarly subject to the first of these effects.

\subsection{Deriving Quenching Timescales}
\label{sec:tquench_method}

\begin{figure}
    \centering
    \includegraphics[width=\linewidth]{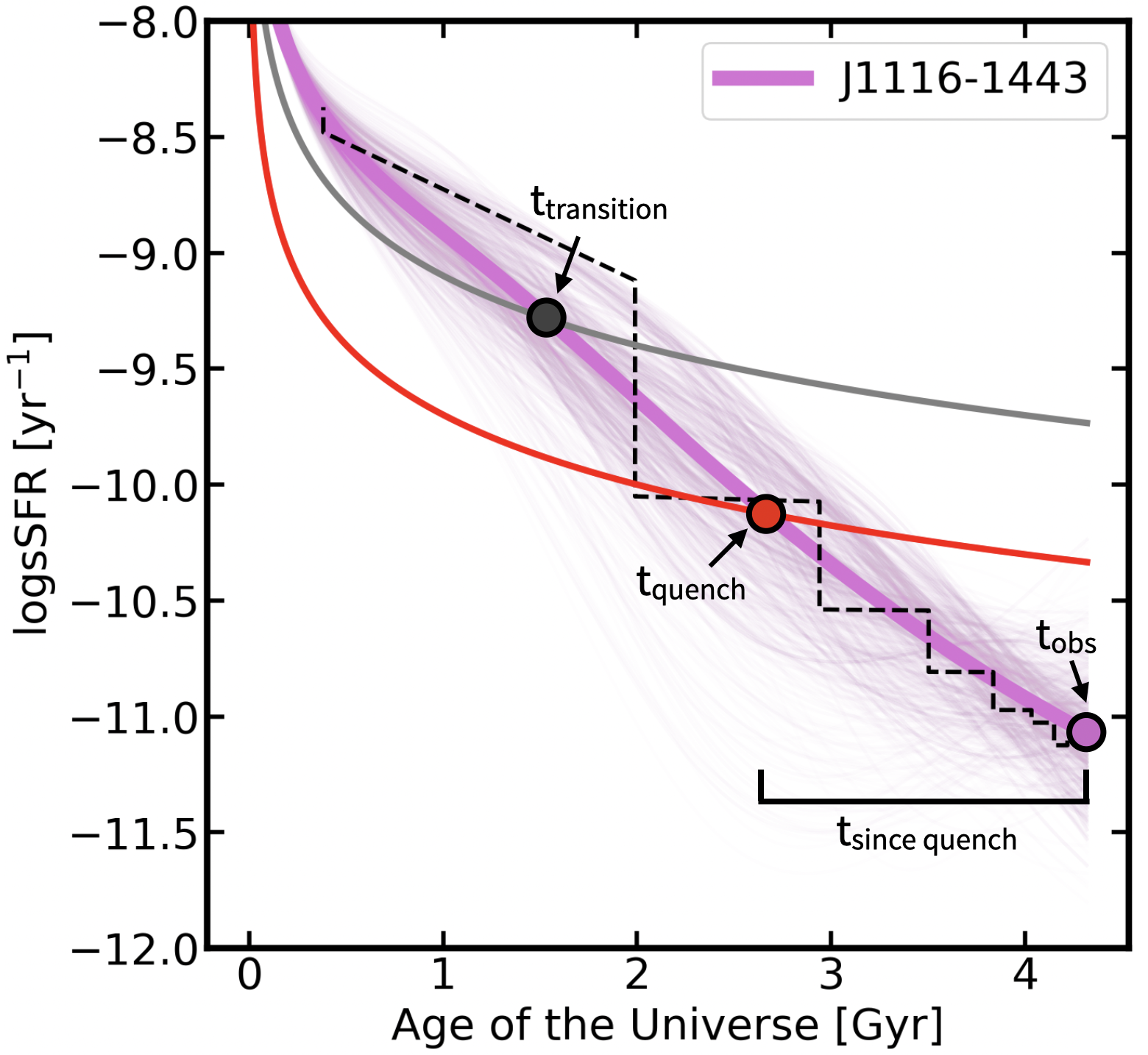}
    \caption{The median smoothed SFH of J1116-1443 is shown (pink line), as well as the SFHs from individual chains and the original 10 bin SFH (black dashed line). The quenching threshold is defined by 0.2/$t_{H}$ (red line), where $t_{H}$ is the age of the Universe; the transition time is defined as 0.6 dex above this relation (gray line). We note where $t_{\mathrm{quench}}$, $t_{\mathrm{transition}}$ and $t_{\mathrm{since \: quench}}$ are located according to these criteria.} 
    \label{fig:sfh}
\end{figure}

Based on the SFHs we obtain from SED fitting (Section \ref{sec:sed}), we derive a quenching timescale and the time since quenching for each galaxy as follows. We smooth the 10 bin SFH by taking the midpoints of each bin and interpolating. We then determine the sSFR at each time step by dividing the SFR by the stellar mass at that time. Integrating the SFH yields the total mass formed, whereas the sSFR shoudl be normalized by the mass that survives, after mass loss from stellar evolution. We therefore correct the integrated mass formed downward by 0.2 dex, corresponding to a surviving mass fraction of $\sim$0.6 for a \citet{chabrier2003} IMF \citep{madua2014}. Although the surviving mass fraction evolves with the age of the stellar population, it declines most rapidly within the first $\sim$1 Gyr after star formation and changes only slowly thereafter. Because $t_{\mathrm{transition}}$ and $t_{\mathrm{quench}}$ occur after the bulk of the stellar mass has formed, a constant correction is adequate for these timescales. It would be less reliable for any timescale derived from the sSFR history at earlier times, when the population is young and the surviving fraction is higher. To determine when a galaxy is quenched, we adopt a criteria of 0.2/$t_H$, where $t_H$ is the age of the Universe \citep{Pacifici2016}. When the sSFR drops below this value, it has quenched; we define this time as $t_{\mathrm{quench}}$. We further define when the sSFR is at 0.6 dex above this cutoff to be the time when the galaxy begins to transition off of the star-forming main sequence (SFMS), or $t_{\mathrm{transition}}$. The time since quenching is then $t_{\mathrm{since \: quench}}=t_{\mathrm{obs}}-t_{\mathrm{quench}}$, or the difference between the time at observation and the time at quenching. Figure \ref{fig:sfh} demonstrates how we derive these times. We also test using the offset from the SFMS measured in \citet{leja2022} as our quenching threshold, and find that it is generally in close agreement to the 0.2/$t_H$ cutoff at $t_{H}>2$ Gyr. However, at earlier times there is a sharp drop in the SFMS likely due to extrapolation; therefore we adopt the 0.2/$t_H$ criteria.

To derive timescale errors, we draw 1000 SFH chains from the posterior distribution and calculate $t_{\mathrm{quench}}$ and $t_{\mathrm{transition}}$ for each chain. We adopt the median, 16th, and 84th percentile when reporting $t_{\mathrm{since \: quench}}$ in Section \ref{sec:discussion}. 

\subsection{Measuring Dust Mass}

\label{sec:dmass}
We estimate dust mass assuming a modified blackbody \citep[e.g.][]{greve2012} with dust emissivity $\beta$ of 1.8 and a dust mass opacity coefficient $\kappa_{353 \text{GHz}}=0.77$ g$^{-1}$cm$^2$ \citep{dunne2000,dacunha2021}. Following \citet{magdis2021}, who measure dust temperatures of stacked quiescent galaxy samples at $0.3<z<2.2$, we adopt a dust temperature of $T_\mathrm{dust}=22.0$K for COOL-LAMPS galaxies at $z\sim1$ and $T_\mathrm{dust}=23.5$ for J1522-2535 at $z\sim2$. For the three detected sources, J1201-3952, J1553-0142, and J2359-0208, we measure the dust continuum flux within apertures derived from modeling optical data with \texttt{GALFIT}. 

We propagate the uncertainty in the dust-to-stellar mass fraction (M$_{\mathrm{dust}}$/M$_{\odot}$) by incorporating errors from SED fitting in determining the stellar mass, in addition to the noise in the dust continuum data used to calculate the dust mass. While the lens models are largely based on ground-based data alone, and therefore relatively uncertain \citep{sharon2020}, we mainly consider relative fractions of physical parameters that largely mitigate lensing uncertainties (e.g., $f_{\mathrm{dust}}$, sSFR). However, our uncertainty analysis does not account for a potentially significant source of systematic uncertainty in our measurement of dust mass, which is dust temperature. If we instead assume dust temperatures range between 15-35K, this could shift the dust-to-stellar mass fraction by $\pm0.3$ dex. Robust constraints on dust temperature require additional data sampling the peak of the far-IR, which are challenging to acquire at cosmic noon for these quiescent systems. Ongoing work performing a joint optical to far-IR modeling of the MRG-M0138 strong lensing system assuming energy balance will start to make progress in this area (Medrano et al. in prep).  To reflect the uncertainty in dust temperature, we add a systematic error bar to all plots that include dust mass.

\section{Results}

\label{sec:results}

\subsection{A Wide Range of Dust to Stellar Mass Ratios}

\begin{figure}
    \centering
    \includegraphics[width=\linewidth]{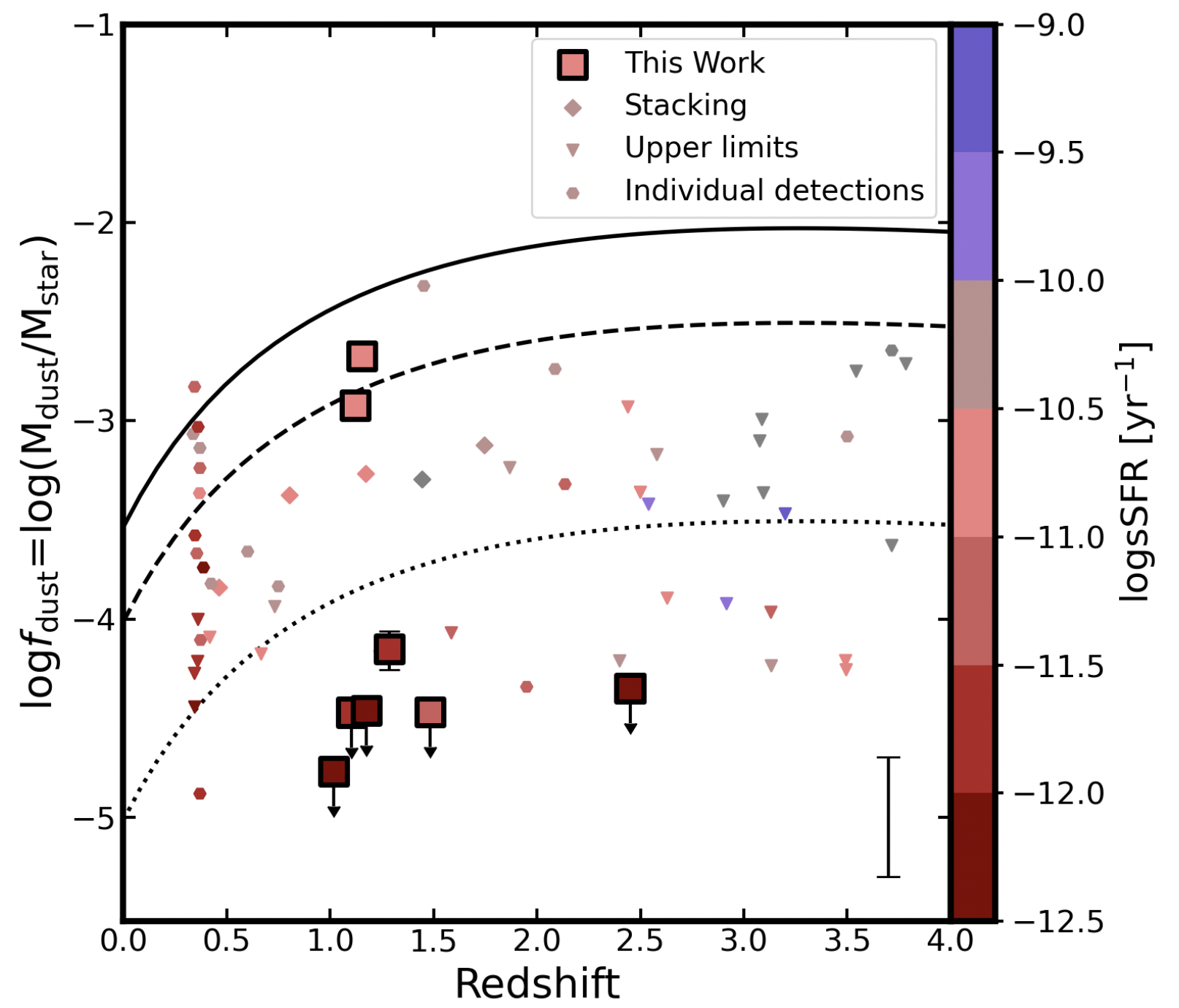}
    \caption{The dust fraction ($f_\mathrm{dust}$=M$_\mathrm{dust}$/M$_\mathrm{star}$) of our sample of quiescent galaxies spans over two orders of magnitude, including objects that are among the dustiest and least dusty quiescent galaxies discovered. Points are color coded by sSFR, with tracks from the \citet{tacconi2018} scaling relation assuming  a gas-to-dust ratio (GDR) of 100 and logM$_{\mathrm{star}}$/ M$_\odot$=11 at the star-forming main sequence (solid line), 0.9 dex below the main sequence with a GDR=100 (dashed line), and 0.9 dex below the main sequence with a GDR=1000 (dotted line). We also include results from the literature \citep{hayashi2018,gobat2018,magdis2021,caliendo2021,kalita2021,whitaker2021,suzuki2022,blanquezsese2023,lee2024,umehata2025,siegel2025,lorenzon2025, spilker2025, chang2026}.We add a systematic error bar (bottom right corner) representing the uncertainty in dust mass introduced by allowing dust temperature to vary from 15-35K.}
    \label{fig:z_dustfrac}. 
\end{figure}

In Figure \ref{fig:z_dustfrac}, we show the ratio of dust mass to stellar mass, $f_\mathrm{dust}$, as a function of redshift, measuring dust mass based on the calculations described in Section \ref{sec:dmass} and stellar mass from SED fitting as described in Section \ref{sec:sed}. As both dust mass and stellar mass are measured in the image plane, the magnification should approximately cancel in $f_{\mathrm{dust}}$.  However, differential magnification could bias $f_{\mathrm{dust}}$ if the dust and stars have different intrinsic distributions within the galaxies, since regions closer to the critical curve are more strongly magnified \citep[e.g.][]{serjeant2012}. The optical and near-IR imaging is seeing-limited ($\sim$0.7--1'') and the ALMA beam is $\sim$1.1''. Hence, the stellar and dust emission are each measured over broadened profiles that average over different parts of the magnification map; at this moment, we cannot directly verify that the two components share the same effective magnifications, or are strictly co-spatial. We compare our sample with other studies of quiescent galaxies, with results from stacking from $z=0.5$ to $z=2$ \citep{gobat2018,magdis2021, blanquezsese2023}, individual measurements at $z\sim$0.4 \citep{lorenzon2025} and $z\sim$0.7 \citep{spilker2025}, and $z>1$ \citep{hayashi2018,caliendo2021, whitaker2021, suzuki2022, lee2024, umehata2025, siegel2025, spilker2025, chang2026}. For reference, we also include the \citet{tacconi2018} scaling relation, assuming a typical molecular gas-to-dust mass ratio, $\delta_{\mathrm{GDR}}$, of 100 and logM$_{\mathrm{star}}$/M$_\odot$=11 for the SFMS, as well as 0.9 dex below the SFMS (the threshold we use to determine time at quenching, see Section \ref{sec:tquench_method}), with a $\delta_{\mathrm{GDR}}$ of 100 and 1000 to account for the wide variety of ratios observed in quiescent galaxies \citep{whitaker2021b,spilker2025,lorenzon2025b}. When assuming $\delta_{\mathrm{GDR}}$=100, most individual measurements of $f_\mathrm{dust}$ from the literature that fall $>$2 dex below the standard \citet{tacconi2018} relation are upper limits, while stacked results and a handful of individual detections are roughly consistent with the relation extrapolated to the defined quenching threshold.

Two of the systems with 1.3mm detections in our sample (J1201-3952 and J2359-0208) have higher $f_\mathrm{dust}$ values than previously measured in lensed quiescent samples at higher redshifts \citep{whitaker2021}, and are more comparable with both the stacked results \citep{magdis2021,gobat2018, blanquezsese2023} and the brightest individual dust detections \citep{hayashi2018, suzuki2022, lee2024} -- however, these earlier studies rely on photometrically-selected quiescent galaxies. The dust masses of these two objects are consistent with the scaling relation between dust mass and redshift found for ALMA-detected star-forming galaxies \citep{donevski2020, kokorev2021}. J1201-3952 in particular may have the highest dust fraction yet measured in a spectroscopically-selected quiescent galaxy, offset 0.3 dex below the SFMS \citet{tacconi2018} relation.  Only the source ALMA.14 \citet{hayashi2018} has a higher dust fraction, but as discussed by \citet{lee2024}, its classification as a quiescent galaxy is based solely on optical/near-infrared photometry, which may significantly underestimate its true SFR; thus its classification as a quiescent galaxy is uncertain.

\begin{figure}
    \centering
    \includegraphics[width=0.9\linewidth]{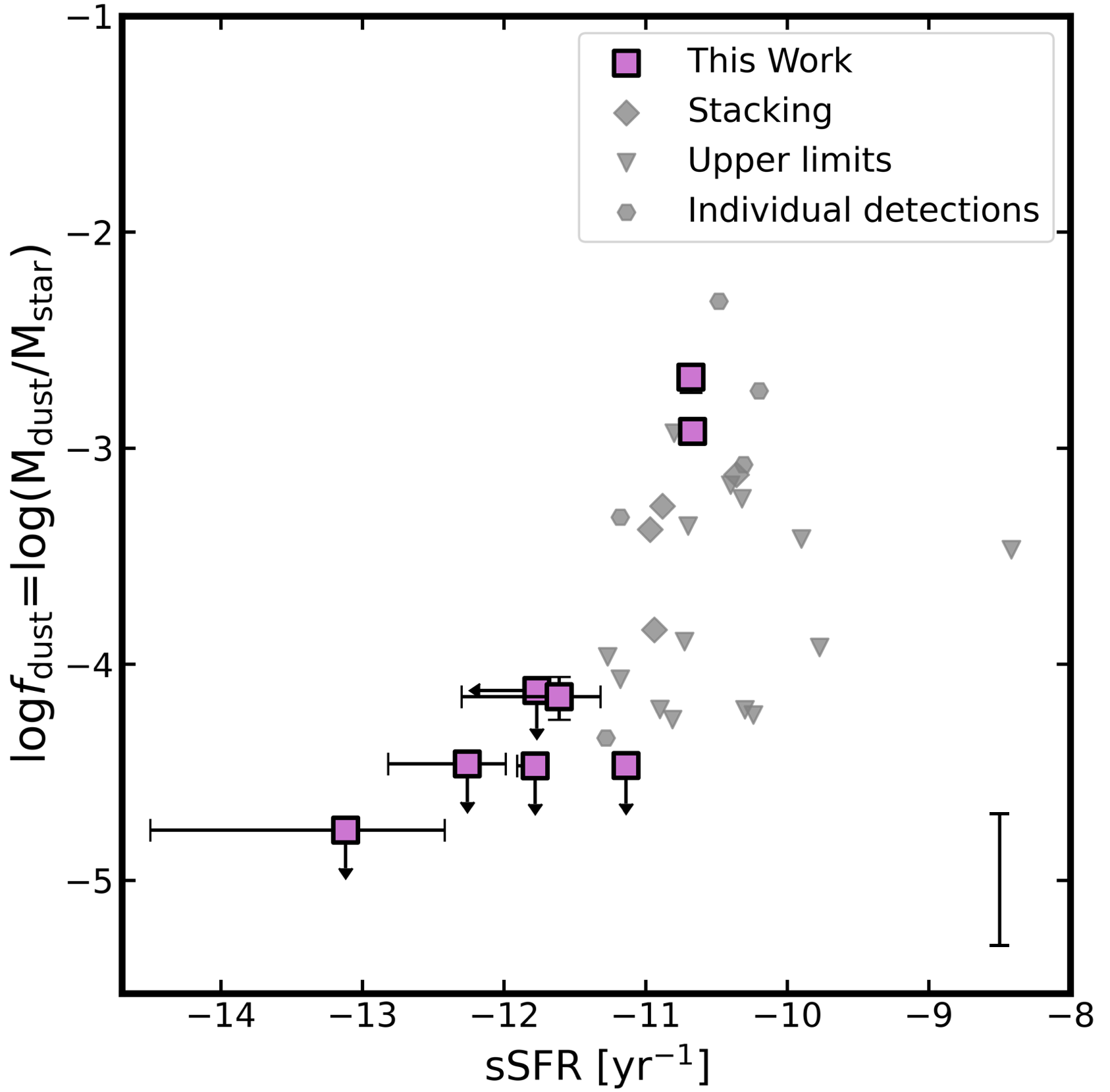}
    \caption{There is a mild association (Kendell $\tau=0.22$) between specific star formation rate (sSFR) and dust fraction ($f_\mathrm{dust}$=M$_\mathrm{dust}$/M$_\mathrm{star}$), and scatter over two order of magnitudes in dust fraction. We add a systematic error bar (bottom right corner) representing the uncertainty in dust mass introduced by allowing dust temperature to vary from 15-35K.}
    \label{fig:sSFR_dustfrac}
\end{figure}

\begin{figure}
    \centering
    \includegraphics[width=\linewidth]{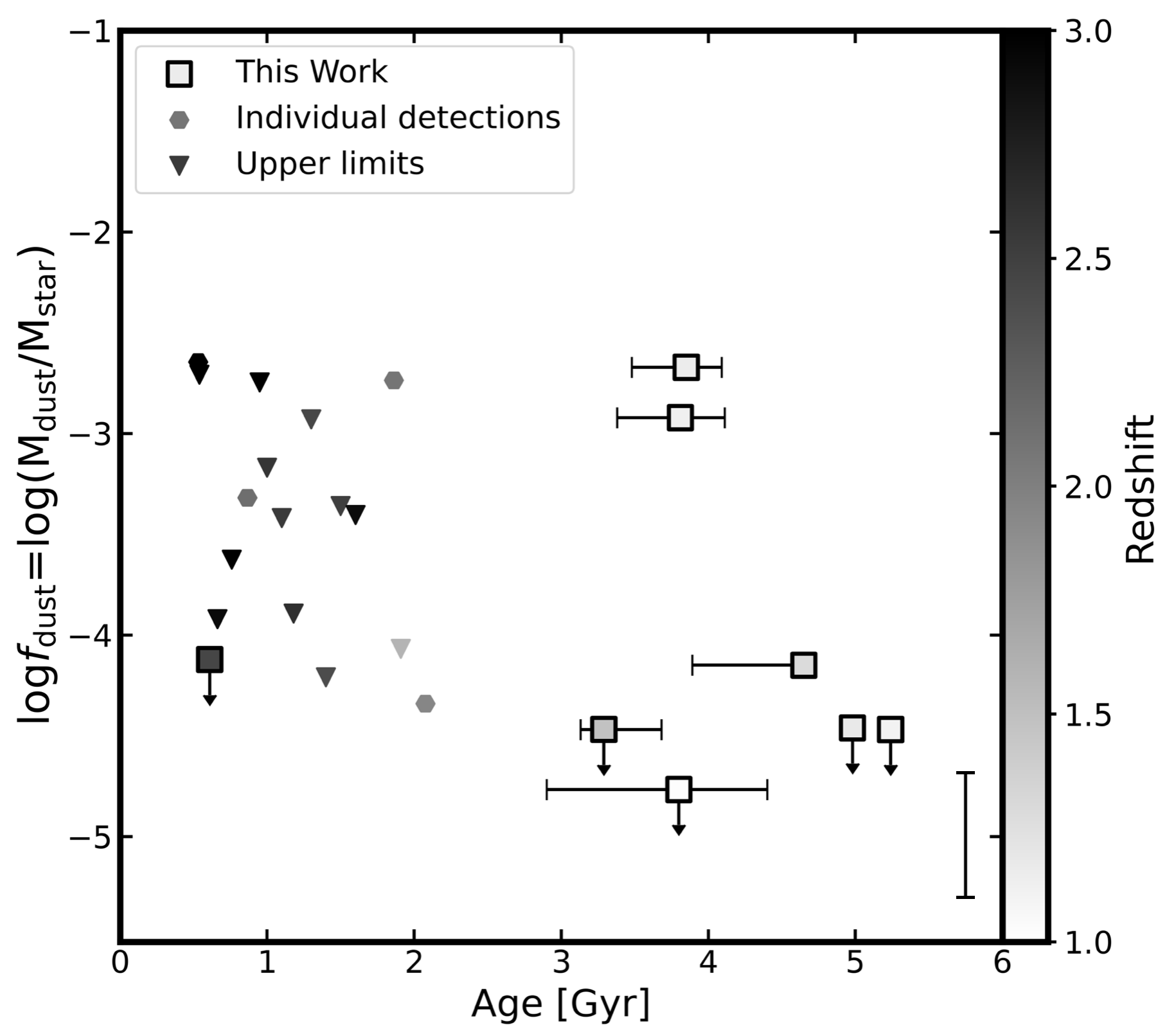}
    \caption{Dust fraction ($f_\mathrm{dust}$=M$_\mathrm{dust}$/M$_\mathrm{star}$) plotted as a function of age; we do not find a statistically significant association between these two properties. Points are color coded by redshift; lower redshift galaxies are older. We add a systematic error bar (bottom right corner) representing the uncertainty in dust mass introduced by allowing dust temperature to vary from 15-35K.}
    \label{fig:age_dustfrac}
\end{figure}

\begin{figure*}
    \centering
    \includegraphics[width=\textwidth]{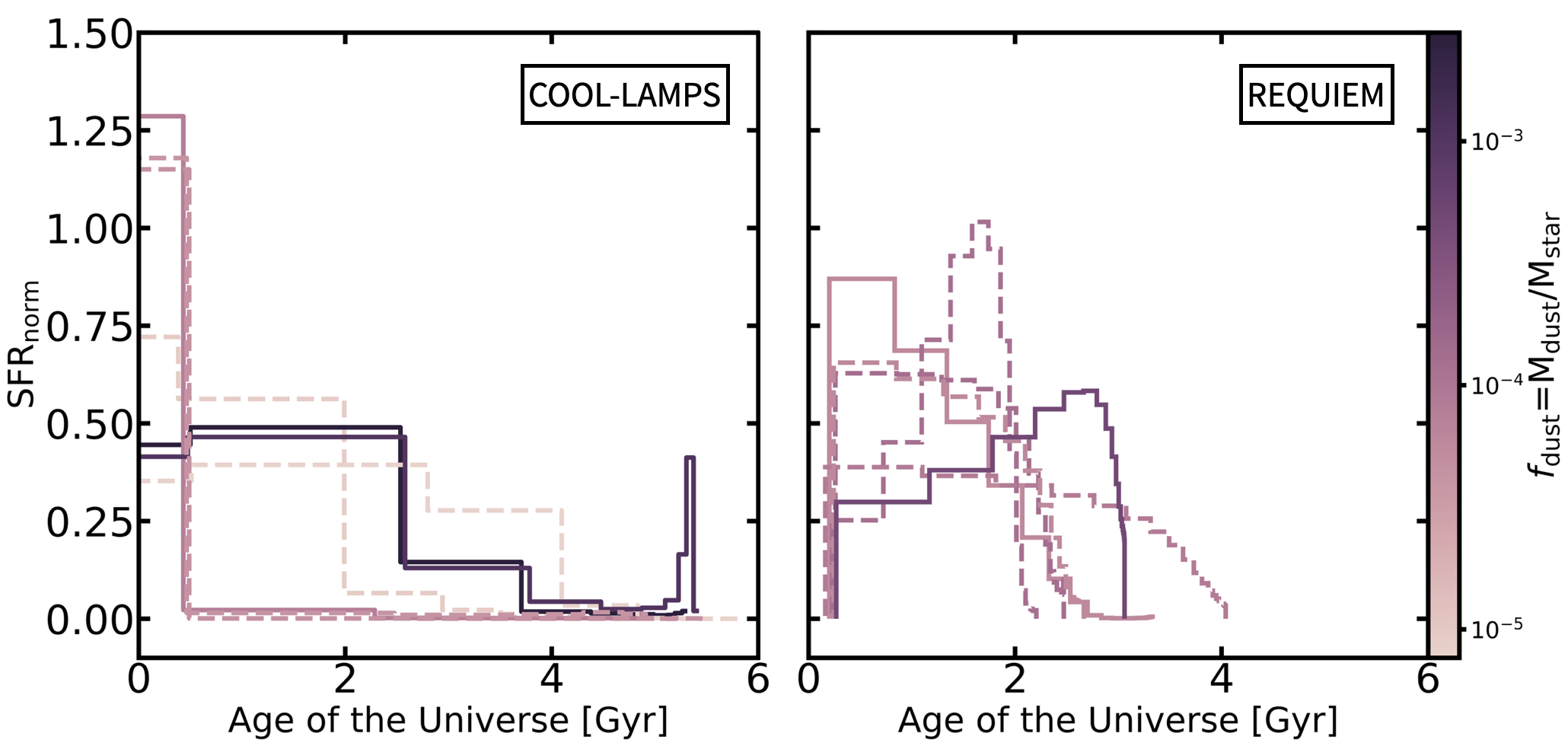}
    \caption{Left: The star formation histories normalized by total mass formed over the entire SFH; the SFRs of several COOL-LAMPS galaxies drop sharply within 2 Gyrs of formation. Two of the galaxies with dust detections (solid lines) show more extended or recent star formation relative to galaxies that are undetected in dust continuum (dashed lines). SFHs are color coded by dust-to-stellar mass fraction; for galaxies with non-detections in dust continuum, the dust fractions are upper limits. Right: The star formation histories normalized by total mass formed for REQUIEM galaxies exhibit a range of shapes, with most undergoing quenching about 2 Gyr after the Big Bang (at $z\sim3$).} 
    \label{fig:sfr_norm}
\end{figure*}

On the other hand, the remainder of the sample is fairly dust poor. J1553-0142 has the second lowest $f_\mathrm{dust}$ value of any quiescent galaxy with a solid dust detection, after MRG-M0138 from the lensed REQUIEM sample \citep{whitaker2021}. Moreover, J1323-0343 and J1116-01443 are the two most stringent upper limits on $f_\mathrm{dust}$, with values of order $f_\mathrm{dust} \sim 2\times 10^{-5}$. In summary, our sample shows a wide diversity in $f_\mathrm{dust}$; additionally, aside from this study -- there are only two other existing dust measurements in individual quiescent galaxies at $1<z<1.5$ \citep{hayashi2018, spilker2025}, so our work provides an 4.5$\times$ increase in the number of such systems.

\subsection{Dust Properties with sSFR and Age}

We plot dust fraction against specific star formation rate (sSFR) in Figure \ref{fig:sSFR_dustfrac}, revealing a mild positive association between $f_\mathrm{dust}$ and sSFR.  We measure the Kendell $\tau$ correlation coefficient using the python package \texttt{pymccorrelation} \citep{privon2020}, treating upper limits as left-censored points \citep{isobe1986}, and find $\tau=0.22$ with a p-value of 0.08. There is significant scatter in sSFR -- at log(sSFR)$\sim -10.5$, the dynamic range of $f_\mathrm{dust}$ values extends over two orders of magnitude. The large scatter in dust fraction may reflect differing timescales in dust destruction, complicated by varying times since quenching across the sample (see Section \ref{sec:tquench}). Two of the galaxies with dust detections at $z\sim1$ (J1201-3952, J2359-0208) have log(sSFR)$\sim$-10.7 yr$^{-1}$, similar to previous studies at $z\sim2$, while the sSFR of J1553-0142 is lower (log(sSFR)$\sim$-11.6 yr$^{-1}$).  The $z\sim1$ galaxies that are undetected tend to have lower values, with log(sSFR)$\lesssim$-11.5 for three out of the four cases.

In Figure \ref{fig:age_dustfrac}, we plot the mass-weighted stellar age as a function of redshift. Stellar ages taken from the literature are mass-weighted \citep{kalita2021, suzuki2022,lee2024, siegel2025}, with the exception of REQUIEM galaxies, which are derived from the time at which 50\% of stellar mass has formed (t$_{50}$)\citep{akhshik2023}. Although it is difficult to surmise a trend given the large number of upper limits, $f_\mathrm{dust}$ appears to generally decrease with increasing stellar population age. We also calculate a Kendell $\tau$ correlation for age and dust fraction, and find $\tau=0.08$ with a p-value of 0.6, indicating a positive association between age and dust fraction with low statistical significance. The lower redshift galaxies in our sample tend to have older ages ($>$3Gyr) relative to higher redshift galaxies ($>$2Gyr).  This largely reflects that higher redshift galaxies have simply had less time to grow their stellar populations and quench, with slower quenching systems only emerging at later times -- high redshift sources are generally caught closer to the epoch of quenching where residual dust is more likely (in the case where quiescent galaxies are the descendants of dusty star-forming galaxies).

\subsection{Star Formation Histories of COOL-LAMPS and REQUIEM Galaxies}

To explore how the dust properties of our sample relate to their SFHs, we reconstruct how the star formation in these galaxies grew and subsequently shut down over cosmic time. In Figure \ref{fig:sfr_norm}, we plot the SFHs normalized by the total mass formed and compare to those derived for the higher redshift REQUIEM galaxies, with stellar mass formed calculated by integrating over the SFH. The details of the SED fitting, from which we derive the SFHs, can be found in Section \ref{sec:sed}.  

For the COOL-LAMPS sample at $z\sim1$ (left, Figure~\ref{fig:sfr_norm}), we find that two of the galaxies with dust detections (2359-0208 and J1201-
3952) have more extended SFHs on average relative to dust-undetected galaxies and J1553-0142. The latter have steeper drops in SFR when the age of the Universe was less than 2 Gyr old. Unlike the rest of the sample, which quench star formation early without rejuvenation, J2359-0208 appears to have experienced a second burst of star formation starting when the Universe was 3 Gyr, with very recent and extreme quenching occurring over the last $\sim$10Myr. We note that the fit for J2359-0208 is uncertain due to a noisy spectrum therefore, the shape of the SFH is more uncertain relative to the rest of our sample.  The uncertainties are folded into our measurements of age, sSFR, and stellar mass.   

Figure~\ref{fig:sfr_norm} (right) shows that the majority of the REQUIEM galaxies (at $z\sim1.5-3$) cross the quenching threshold when the Universe was $\sim$2 Gyr, close to the epoch of observation, with no evidence of residual cold dust remaining, though for a range of quenching timescales \citep{akhshik2023}. The most extended quenching timescale (MRG-M0138), more gradual than even the lower redshift COOL-LAMPS galaxies, has a tightly constrained dust detection at $f_\mathrm{dust}\sim6\times10^{-5}$, with the dust residing only in the core \citep{whitaker2021}. Similarly, the other REQUIEM galaxy with a dust detection, MRG-M2129, shows centrally-concentrated dust at a more moderate $f_{\mathrm{dust}}$ value, with the overall (s)SFR similarly elevated relative to the other galaxies in the sample.  This stands in contrast to the COOL-LAMPS galaxies J2359-0208 and J1201-3952, where the cold dust is unambiguously extended and more closely follows the stellar light profile, which we discuss in more detail in Section~\ref{sec:discussion}.

The REQUIEM and COOL-LAMPS samples seem to collectively offer us a glimpse into the life cycle of dust post-quenching and an opportunity to empirically constrain dust destruction conditions and associated timescales.  Figure~\ref{fig:tquench} (left) shows the full sample, with the measurements of dust fraction relative to the time that has elapsed since crossing the quenching threshold for each of these systems (derived according to the methodology described in Section \ref{sec:tquench_method}). As one can also gather from Figure~\ref{fig:sfr_norm}, the REQUIEM galaxies have all recently shut down star formation within the last Gyr, and often only a few hundred Myrs before the epoch of observation.  Conversely, the COOL-LAMPS sample spans a wide range of time since quenching and a similarly wide range of dust fractions. While this sample is notably biased due to the nature of strong-lens system discovery, the combined set of REQUIEM and COOL-LAMPS galaxies broadens the parameter space we can explore across redshift, SFH shapes, and dust content, yielding some general trends that we discuss next.

\section{Discussion}
\label{sec:discussion}

\begin{figure*}
    \centering
    \includegraphics[width=\textwidth]{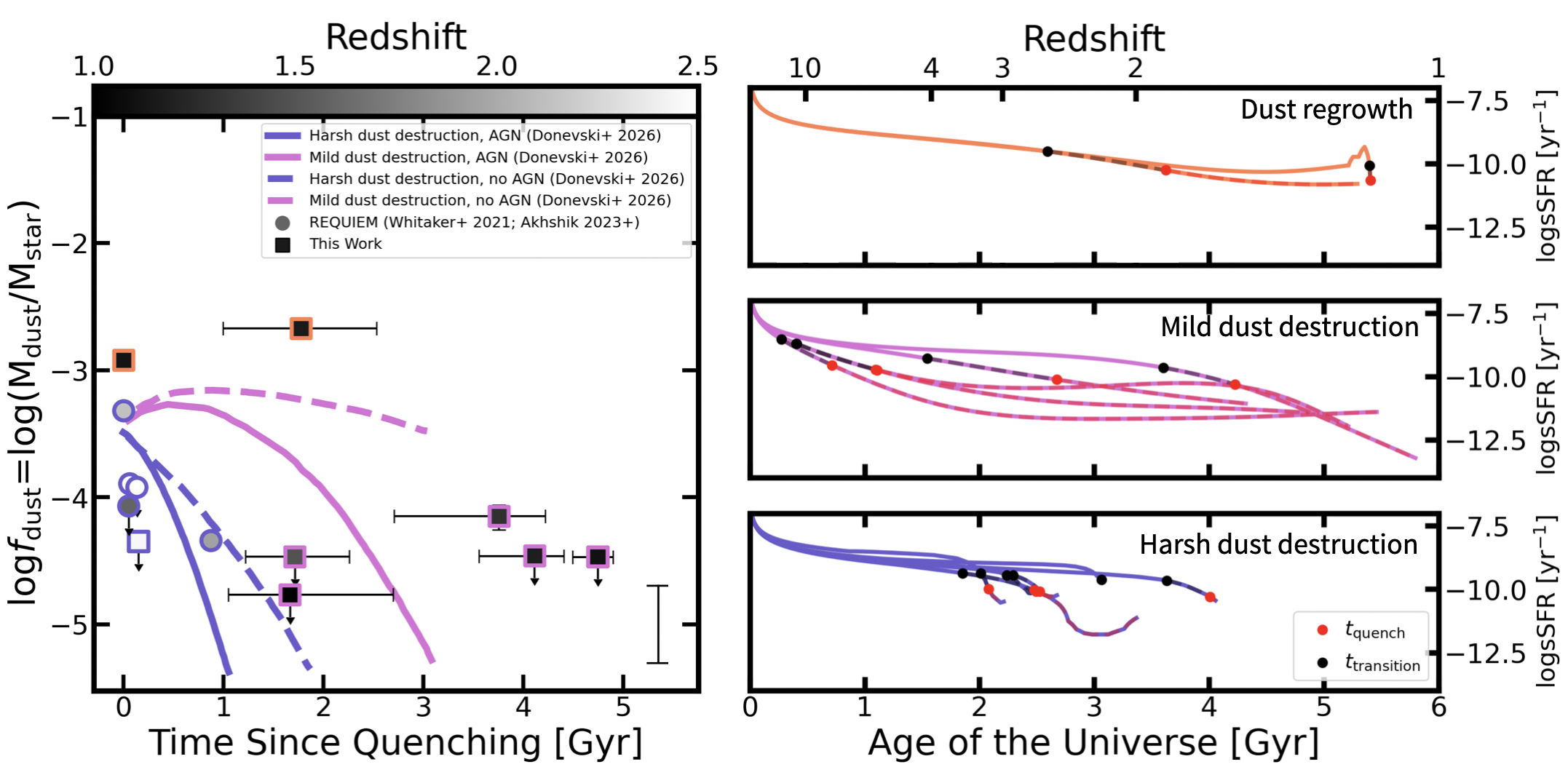}
    \caption{Left: Lower redshift COOL-LAMPS galaxies quenched further in the past, and have dust fractions consistent with mild dust destruction or dust regrowth. Higher redshift REQUIEM galaxies are more recently quenched and are more consistent with harsh dust destruction. We add a systematic error bar (bottom right corner) representing the uncertainty in dust mass introduced by allowing dust temperature to vary from 15-35K. Right: The star formation histories for COOL-LAMPS and REQUIEM galaxies, grouped by where they fall on the time since quenching vs dust fraction plot. SFHs with dust regrowth/mild dust destruction are shown at the top, mild dust destruction in the middle, and harsh dust destruction on the bottom.}
    \label{fig:tquench}
\end{figure*}

It has been predicted that dust should be destroyed over rapid timescales after quenching \citep{li2019}, although recent models suggest that this is highly dependent on the density and energetics of the local ISM, with dust regrowth possible in cold, dense, and less energetically disturbed environments \citep{lorenzon2025, donevski2026}. In Figure \ref{fig:tquench}, we compare our observed dust fractions as a function of time since quenching to the model predictions of \citet{donevski2026}. The solid lines represent dust destruction trajectories when AGN feedback is included, whereas the dotted lines do not include AGN feedback.  Turning on AGN feedback is associated with increased dust destruction (decreasing $f_{\mathrm{dust}}$) with increasing severity as more time elapses post-quenching.  In effect, the AGN energetics likely keep the gas within the extended dark matter halos hot for prolonged periods, with thermal sputtering from hot electrons destroying the dust \citep{whitaker2021b}. This destruction is irrevocable in the case where it is coupled with similarly harsh ISM conditions, though can extend over several gigayears for more mild ISM conditions \citep{donevski2026}.

\subsection{Changing Modes of Dust Destruction with Redshift}
\label{sec:tquench}

We find that the higher redshift galaxies from the REQUIEM sample have quenched more recently, closer to the epoch of observation, and show dust fractions that align with harsh dust destruction, likely driven at least in part by AGN feedback. This is consistent with recent JWST observations that reveal neutral outflows from AGN in recently quenched systems at $z>2$ \citep{belli2024, davies2024, bugiani2025, luo2026, zhu2026}, indicating the prevalence of AGN feedback at earlier times. For galaxies in this sample with dust detections, the dust is compact and centrally located, which may be the result of recent (in the last several hundred Myr) star formation in the core. While MRG-M2129 has a moderate dust fraction, its SFH is consistent with recent, rapid quenching. We therefore place it in harsh-destruction category, as it is observed shortly after quenching, before its residual core dust has had time to be destroyed. There is evidence that MRG-M2129 hosts an AGN \citep{newman2018, man2021, morishita2022}, supporting that AGN feedback could play a role in the process of dust destruction.

By contrast, the lower redshift COOL-LAMPS galaxies are largely consistent with a mild dust destruction scenario at face value. That said, four COOL-LAMPS systems (J0112-2053, J1116-1443, J1323-0343, and J1509-3308) have stringent upper limits on $f_{\mathrm{dust}}$ that do not rule out more harsh ISM conditions, similar to the REQUIEM sample.  These sources also show similar SFHs, with early formation and rapid quenching, observed $>$3 Gyr after quenching. By contrast, the two COOL-LAMPS systems with extended dust emission are more unambiguously overlapping with the model tracks for mild ISM conditions and dust regrowth.  The difference in the implied dust destruction timescales may reflect the changing ISM conditions and a shift in the dominant physical mechanisms driving quenching \citep[see][for a review]{whitaker2026}. At $z\sim2$, more intense feedback leads to more efficient dust destruction, while these processes become slower by $z\sim1$. 

\subsection{Grain Regrowth and Dust Production after Quenching}

In the cases of J1201-3952 and J2359-0208,  their dust fractions exceed even the most optimistic dust destruction scenario with no AGN feedback. J2359-0208 has a SFH consistent with a post-starburst classification, with a relatively high sSFR that is near the quenching boundary (e.g. the sSFR is $\sim$0.9 dex below the SFMS), similar to MRG-M2129, although J2359-0208 also experienced a late rejuvenation followed by rapid quenching close to the epoch of observation (Figure~\ref{fig:sfr_norm}. While both do not formally qualify as quiescent according to the criteria from \ref{sec:tquench_method}, and as such we assign both a time since quenching of zero, their dust morphologies show a striking difference:  MRG-M2129 retains compact dust in its core \citep{whitaker2021, morishita2022}, consistent with the harsh, AGN-driven destruction discussed in Section~\ref{sec:tquench}, whereas J2359-0208 shows extended dust emission. J2359-0208 may appear consistent with dust growth because rejuvenation and stellar dust production combine, with fresh dust from the recent burst adding to the grain regrowth and/or TP-AGB dust production by the intermediate-age stellar population formed during rejuvenation.  For the latter case, dust from TP-AGB stars should trace the spatial distribution of the stellar population formed $\sim$0.2--2 Gyr before observation, here the rejuvenation episode, rather than the overall stellar light. Residual dust that has not yet had enough time to deplete, on the other hand, would primarily be expected to reside in the core where higher gas densities shield grains and slow the destruction timescale \citep{richie2026}. Distinguishing the dust distributions robustly will require higher-resolution continuum imaging than the present data provide, together with resolved stellar-population maps.

The dust excess in J1201-3952 and J2359-0208 relative to even the mildest destruction tracks may also point to grain growth in the ISM that is more efficient than in the fiducial models, for example if metals re-accrete onto grains on timescales shorter than $\sim$100 Myr \citep{donevski2026}. Rejuvenation, as observed in a minority of quiescent systems \citep[e.g.,][]{akhshik2021,chauke2019} and predicted by simulations \citep{remus2025,kimmig2025,fortune2025}, provides a complementary channel, supplying fresh dust both directly and by enabling grain growth \citep{lorenzon2025}. However, both interpretations rely on the assumption that the far-IR emission is not powered by ongoing star formation. 

\subsection{Are Dust-Rich Systems Truly Quiescent?}

An obvious concern is whether these dust-rich systems are truly quiescent, or whether they host dust-obscured star formation that is invisible in the rest-frame optical spectra used to derive their SFHs. Strong Balmer absorption and weak emission lines do not uniquely identify a quenched galaxy, as an obscured starburst can sit alongside, or be spatially offset from, an optically post-starburst stellar population \citep{schreiber2018}. For example, among local E+A galaxies with molecular gas measurements, a subset (7 of 26) host obscured starbursts detectable only in the far-IR, some with starburst-like SFRs \citep{baron2023}, although infrared luminosities alone can substantially overestimate SFRs in galaxies below the main sequence \citep{utomo2014,wild2025}. In the FIREbox simulations, bursty, interaction-driven SFHs produce gas-rich post-starburst ``impostors'' that are still forming stars at rates typical of star-forming galaxies \citep{cenci2026}. 

However, a substantial dust reservoir, as we see in J1201-3952 and J2359-0208, does not by itself imply star formation. Local post-starbursts host compact, turbulent molecular reservoirs that form stars roughly an order of magnitude less efficiently than starbursts of similar gas surface density \citep{smercina2022}. Furthermore, at $z\sim0.7$, the IR SEDs of most post-starbursts can be reproduced by dust heated by their evolved stellar populations without buried star formation \citep{setton2025}, although spatially resolved Pa$\alpha$ shows that the most CO-luminous systems are still forming stars behind optically thick dust \citep{setton2026}. Our ALMA Band~6 continuum samples the Rayleigh--Jeans tail and therefore constrains dust mass rather than the infrared luminosity; it cannot by itself rule out obscured star formation, since $L_{\mathrm{IR}}$ inferred from a single long-wavelength point depends steeply on the assumed dust temperature. 

Two lines of evidence nonetheless argue against a hidden starburst. First, the dust fractions of J1201-3952 and J2359-0208 remain 0.3-0.6 dex below those of main-sequence galaxies at similar stellar mass and redshift, when assuming a value of $\delta_{\mathrm{GDR}}$ of 100 (Figure~\ref{fig:z_dustfrac}), whereas an obscured starburst sufficient to place them on the main sequence would typically be accompanied by a dust reservoir comparable to those of star-forming galaxies. Second, the dust emission in both systems is extended and broadly co-spatial with the stars, rather than arising from a compact nuclear or spatially offset clump as seen in misclassified ``quiescent'' galaxies \citep{schreiber2018}. We note, however, that at the resolution of our ALMA data ($\sim$1-1.5$''$), a compact obscured component contributing up to 39\%  and 28\% of the continuum flux could remain unresolved within the extended emission for  J1201-3952 and J2359-0208 respectively \citep[see also][]{sun2022}, so this argument constrains only an obscured starburst that dominates the dust emission.

We therefore cannot rule out modest obscured star formation in these two systems. If present, it would place them in a still-quenching or rejuvenating phase rather than true quiescence, consistent with the rejuvenation already seen in the SFH of J2359-0208. The dust-poor systems, however, are unaffected, as their ultra-deep continuum limits themselves make significant obscured star formation unlikely. Multi-band ALMA continuum spanning the Rayleigh--Jeans tail and approaching the far-IR peak would jointly constrain $T_{\mathrm{dust}}$ and the emissivity index $\beta$, removing the dominant uncertainty in converting the far-IR emission to either a dust mass or an obscured SFR \citep{cochrane2022}. Independently, rest-frame near-IR hydrogen recombination lines (e.g., Pa$\alpha$, Pa$\beta$), which are far less sensitive to attenuation \citep[e.g.,][]{setton2026}, and mid-IR PAH emission would directly test for obscured star formation, though PAHs in quiescent systems can also be excited by evolved stellar populations.

\subsection{Dust Yields from AGB stars and SN}

\begin{figure}
    \centering
    \includegraphics[width=\linewidth]{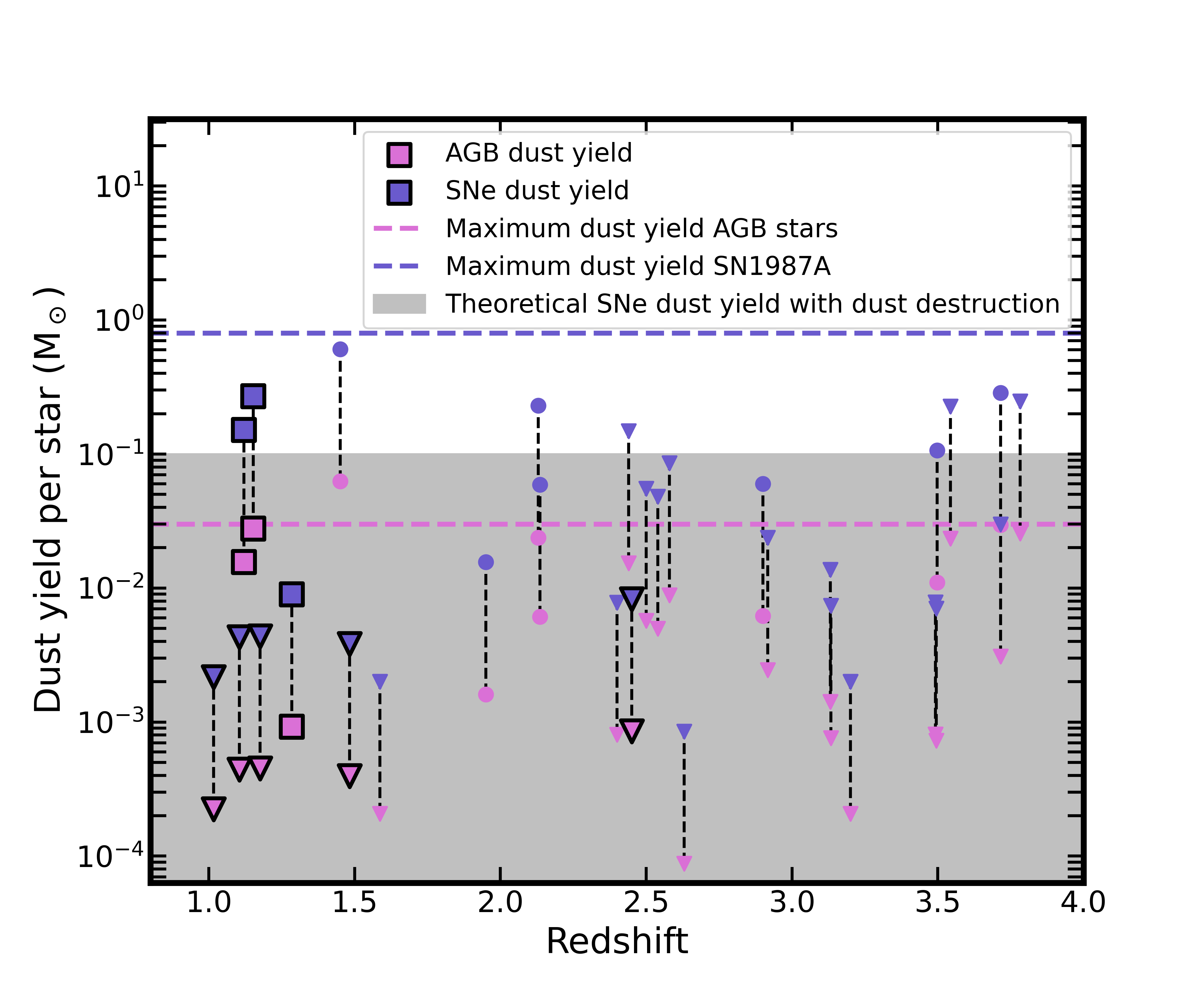}
    \caption{The dust yield per star for either AGB stars or Type II SN to account for the total dust mass approaches or exceeds the theoretical limit for several $z>1$ quiescent galaxies. Results from this study are outlined in black.} 
    \label{fig:dust_yield}
\end{figure}

Taking the quiescent interpretation at face value, we quantify the dust yield from AGB stars and Type II supernovae (SN) required to account for the observed dust masses, following \citet{michalowski2010} and \citet{lee2024}. We assume a Salpeter IMF \citep{salpeter1955} and that stellar mass ranges from $M_{\mathrm{star}}=8-40$M$_\odot$ for Type II SNe and $M_{\mathrm{star}}=1.5-8$M$_\odot$ for AGB stars. \citet{ventura2012} and \citet{nanni2013} report maximum dust yield for AGB stars $\sim$0.25 M$_\odot$, although estimated dust yields vary significantly with estimated initial mass and metallicity; we adopt a maximum dust yield of 0.03M$_\odot$ per AGB star. We adopt a maximum dust yield for Type II SN of 0.8 M$_\odot$ from SN1987A \citep[dust mass measurements vary for this SN from 0.23-0.8 M$_\odot$, see ][]{Indebetouw2014, dwek2015, matsuura2015}, although once sufficient time has passed for the reverse shock to destroy dust grains, it is predicted the dust yield from Type II SN should drop to $<0.1$M$_\odot$ \citep{bocchio2016}. Figure \ref{fig:dust_yield} shows the dust yield per star assuming that all of the dust mass comes from either Type II SN or AGB stars. In most cases, the dust mass that we observe could be explained entirely by either Type II SN or AGB stars. However, the two dust-rich sources, J2359-0208 and J1201-3952, approach or exceed the theoretical limits when attributing the dust to one mechanism alone. In particular, J1201-3952 falls on the 0.03M$_\odot$ limit for AGB stars and sits higher than models that predict a dust yield of $<0.1$M$_\odot$ for Type II SN when including dust destruction from SN shocks. These dust yield estimates therefore suggest efficient post-quenching grain regrowth is possible in 2 out of the 11 systems ($\sim$20\%). This is lower than the $\sim$50\% of simulated quiescent galaxies that maintain roughly constant dust content for $\sim$2 Gyr after quenching in SIMBA \citep{lorenzon2025}, but is consistent with their prediction that such systems are more common at lower stellar masses than those investigated in this paper \citep[see also][]{donevski2026}. 

The dust yield calculation treats each production channel as integrated over the full stellar population, but the rate of AGB dust injection at the epoch of observation depends on the shape of the SFH. Stars of 1.5--8 $M_\odot$ reach the AGB between $\sim$40 Myr and $\sim$2--3 Gyr after they form, and the thermally pulsing AGB phase dominates the near-IR light of stellar populations aged $\sim$0.2--2 Gyr \citep{maraston2005}; since these same stars are efficient dust producers \citep[e.g.,][]{nanni2013, nanni2018}, AGB dust injection should likewise be strongest at intermediate ages. The SFHs in Figure~\ref{fig:sfr_norm} show a qualitative trend in this direction. The dust-poor COOL-LAMPS systems formed nearly all of their stellar mass in a single burst within the first $\sim$0.5 Gyr, and are observed several gigayears later, when the stars now reaching the AGB fall near or below the $\sim$1.5 $M_\odot$ threshold for significant dust production. By contrast, the two dust-rich COOL-LAMPS systems formed their stars over an extended period of several gigayears, with low-level star formation continuing close to the epoch of observation (through a late rejuvenation in J2359-0208, and through a prolonged decline along the quiescence threshold in J1201-3952). Both therefore host stellar populations that are still feeding the TP-AGB phase. The most dust-rich REQUIEM system likewise formed a large fraction of its mass within the last $\sim$1 Gyr before observation. In the REQUIEM sample, however, recent star formation and a short time since quenching coincide, so residual dust and fresh AGB dust cannot be separated. The COOL-LAMPS systems break this degeneracy, with J1201-3952 quenched $\sim$1.8 Gyr before observation yet has the highest $f_{\mathrm{dust}}$ in the sample, suggesting that its extended SFH, rather than a recent quenching event, sustains its dust reservoir. Quantifying this trend requires SFHs with sub-gigayear precision at these ages, which can be tested further with future rest-frame near- to mid-IR spectroscopy.

\section{Conclusion}
\label{sec:conclusion}

In this paper, we combine new ALMA Band 6 dust continuum observations of eight strongly lensed, spectroscopically-selected quiescent galaxies at $z$$\sim$1-2.5 from the COOL-LAMPS and REQUIEM surveys with five archival REQUIEM measurements \citep{whitaker2021,akhshik2023}.  This sample is the largest set of individually targeted quiescent galaxies at cosmic noon with both deep dust continuum constraints and spectroscopically derived SFHs. Our key findings are:

\begin{enumerate}
    \item The dust-to-stellar mass ratio spans more than two orders of magnitude, from $f_{\mathrm{dust}}<10^{-5}$ to $>10^{-3}$, including some of the most stringent upper limits to date and J1201-3952, which may have the highest $f_{\mathrm{dust}}$ yet measured in a spectroscopically-selected quiescent galaxy. Our sample increases the number of individual dust measurements in quiescent galaxies at $1<z<1.5$ by a factor of 4.5.
    \item Compared with the models of \citet{donevski2026}, the recently quenched $z\sim2$ REQUIEM galaxies and the early-forming, rapidly quenched $z\sim1$ COOL-LAMPS galaxies are consistent with harsh dust destruction, likely driven by AGN feedback, whereas the more slowly quenched COOL-LAMPS systems with extended dust are consistent with mild destruction or dust regrowth. This suggests that dust survival depends on how galaxies quench, and possibly that the dominant quenching mode evolves between $z\sim2$ and $z\sim1$.
    \item The two most dust-rich galaxies, J1201-3952 and J2359-0208, exceed even the mildest model destruction tracks, and their dust masses approach or exceed the maximum yields from AGB stars or Type II supernovae alone. Their extended SFHs, including a late rejuvenation in J2359-0208, point to efficient ISM grain growth and/or ongoing TP-AGB dust production by intermediate-age stellar populations.
    \item These interpretations carry two main caveats: the unknown dust temperature introduces a $\pm$0.3 dex systematic uncertainty in $f_{\mathrm{dust}}$, and our single-band, $\sim$1$''$ continuum cannot exclude modest obscured star formation in the two dust-rich systems.
\end{enumerate}

While calibrations of dust continuum as a molecular gas tracer remain uncertain in quiescent galaxies \citep{spilker2025}, this study illustrates the value of dust continuum for studying quenching and feedback. The dust content of quiescent galaxies at cosmic noon varies by more than two orders of magnitude, consistent with a picture in which the mode of quenching (slow versus rapid) and the strength of feedback (mild versus harsh) have lasting consequences for their dust. Three observational steps would test this picture most directly. First, the multi-band and higher-resolution ALMA continuum described in Section~\ref{sec:discussion} would both constrain dust masses and separate compact residual dust from extended, stellar-produced dust. Second, JWST rest-frame mid-IR spectroscopy would directly trace circumstellar dust around AGB stars, testing the contribution of late-stage stellar dust production \citep[e.g.,][]{donevski2026,setton2026}. Third, CO observations would test whether dust traces the remaining molecular gas, given evidence that $\delta_{\mathrm{GDR}}$ in quiescent galaxies can be unusually high \citep{spilker2025}. 

Strong lensing currently makes these measurements feasible for individual galaxies, bringing intrinsically faint dust continuum within reach of ALMA and beginning to resolve the dust relative to the stars, the key discriminant between residual core dust and dust produced by the stellar populations. Looking further ahead, the recently selected PRobe far-Infrared Mission for Astrophysics \citep[PRIMA;][]{prima} will sample the far-IR peak of cold dust emission for statistical samples of quiescent galaxies, making it possible to map how dust survival depends on quenching pathway across the population rather than galaxy by galaxy.

\section{Acknowledgements}

The University of Massachusetts Amherst acknowledges that it was founded and built on the unceded homelands of the Pocumtuc Nation on the land of the Norrwutuck community. We recognize and honor the ongoing presence and stewardship of Indigenous peoples. As a Land Grant institution, UMass Amherst also occupies land obtained through the Morrill Act of 1862, which expropriated Indigenous territories to support the development of public universities. We recognize this historical context and the responsibilities it entails. 

We also wish to acknowledge the many individuals whose labor sustains the institutions that make this research possible,
including the janitorial, facility maintenance, technical, and
administrative support staff at the University of Massachusetts,
Amherst, whose contributions are often unrecognized but essential
to the functioning of the scientific enterprise. 

SB thanks Samantha Berek for helpful discussion. SB and KEW acknowledge support from the National Science Foundation under CAREER Award No. AST-2236773. GK would like to thank the Baum Grant and Fellowship at the University of Washington for support during this work, as well as the ALMA Ambassador Program (administered by NAASC and NRAO). GK acknowledges the International Space Science Institute (ISSI), Bern, for their hospitality, financial support and collaboration during the time of writing this manuscript, as part of their Visiting Scientist program. GK would also like to thank the DiRAC Institute in the Department of Astronomy at the University of Washington. The DiRAC Institute is supported through generous gifts from the Charles and Lisa Simonyi Fund for Arts and Sciences, Janet and Lloyd Frink, and the Washington Research Foundation. This research was supported in part by the University of Pittsburgh Center for Research Computing, RRID:SCR\_022735, through the resources provided. Specifically, this work used the H2P/MPI cluster, which is supported by NSF award number OAC-2117681. JAD acknowledges funding from the Dunlap Institute,  funded through an endowment established by the David Dunlap family and the University of Toronto

A subset of spectra in the COOL-LAMPS program are observations made with the Nordic Optical Telescope, owned in collaboration by the University of Turku and Aarhus University, and operated jointly by Aarhus University, the University of Turku and the University of Oslo, representing Denmark, Finland and Norway, the University of Iceland and Stockholm University at the Observatorio del Roque de los Muchachos, La Palma, Spain, of the Instituto de Astrofisica de Canarias. The NOT data were obtained under program ID P61-031. The data presented here were obtained [in part] with ALFOSC, which is provided by the Instituto de Astrofisica de Andalucia (IAA) under a joint agreement with the University of Copenhagen and NOT.

\bibliography{bib}{}

@ARTICLE{Adscheid2025,
       author = {{Adscheid}, Sylvia and {Magnelli}, Benjamin and {Ciesla}, Laure and {Liu}, Daizhong and {Schinnerer}, Eva and {Bertoldi}, Frank},
        title = "{A$^3$COSMOS: The dust content of massive quiescent galaxies and its evolution with cosmic time}",
      journal = {arXiv e-prints},
         year = 2025,
        month = aug,
          eid = {arXiv:2508.18097},
        pages = {arXiv:2508.18097},
          doi = {10.48550/arXiv.2508.18097},
archivePrefix = {arXiv},
       eprint = {2508.18097},
 primaryClass = {astro-ph.GA},
       adsurl = {https://ui.adsabs.harvard.edu/abs/2025arXiv250818097A}
}

@ARTICLE{suyu2026,
   author = {{Suyu}, S.~H. and {Acebron}, A. and {Grillo}, C. and
             {Bergamini}, P. and {Caminha}, G.~B. and {Cha}, S. and
             {Diego}, J.~M. and {Ertl}, S. and others},
    title = "{Cosmology with supernova Encore in the strong lensing cluster
             MACS J0138$-$2155: Lens model comparison and H$_0$ measurement}",
  journal = {\aap},
     year = 2026,
    month = apr,
   volume = {708},
      eid = {A291},
      doi = {10.1051/0004-6361/202557235},
archivePrefix = {arXiv},
   eprint = {2509.12319}
}

@ARTICLE{acebron2025,
   author = {{Acebron}, A. and {Bergamini}, P. and {Rosati}, P. and others},
    title = "{Enhanced strong-lensing model of MACS J0138.0$-$2155 based on
             new JWST and VLT/MUSE observations}",
  journal = {\aap},
     year = 2025,
    month = jul,
   volume = {699},
      eid = {A101},
      doi = {10.1051/0004-6361/202554468},
archivePrefix = {arXiv},
   eprint = {2503.09694}
}

@ARTICLE{ertl2025,
   author = {{Ertl}, S. and {Suyu}, S.~H. and others},
    title = "{Cosmology with supernova Encore in the strong lensing cluster
             MACS J0138$-$2155: Photometry, cluster members, and lens mass
             model}",
  journal = {\aap},
     year = 2025,
    month = oct,
   volume = {702},
      eid = {A157},
      doi = {10.1051/0004-6361/202554477},
archivePrefix = {arXiv},
   eprint = {2503.09718}
}

@ARTICLE{rodney2021,
   author = {{Rodney}, S.~A. and {Brammer}, G.~B. and {Pierel}, J.~D.~R.
             and others},
    title = "{A gravitationally lensed supernova with an observable
             two-decade time delay}",
  journal = {Nature Astronomy},
     year = 2021,
   volume = {5},
    pages = {1118-1125},
      doi = {10.1038/s41550-021-01450-9}
}

@ARTICLE{Dey_2019,
       author = {{Dey}, Arjun and {Schlegel}, David J. and {Lang}, Dustin and
         {Blum}, Robert and {Burleigh}, Kaylan and {Fan}, Xiaohui and
         {Findlay}, Joseph R. and {Finkbeiner}, Doug and {Herrera}, David and
         {Juneau}, St{\'e}phanie and {Landriau}, Martin and {Levi}, Michael and
         {McGreer}, Ian and {Meisner}, Aaron and {Myers}, Adam D. and
         {Moustakas}, John and {Nugent}, Peter and {Patej}, Anna and
         {Schlafly}, Edward F. and {Walker}, Alistair R. and
         {Valdes}, Francisco and {Weaver}, Benjamin A. and
         {Y{\`e}che}, Christophe and {Zou}, Hu and {Zhou}, Xu and
         {Abareshi}, Behzad and {Abbott}, T.~M.~C. and {Abolfathi}, Bela and
         {Aguilera}, C. and {Alam}, Shadab and {Allen}, Lori and {Alvarez}, A. and
         {Annis}, James and {Ansarinejad}, Behzad and {Aubert}, Marie and
         {Beechert}, Jacqueline and {Bell}, Eric F. and {BenZvi}, Segev Y. and
         {Beutler}, Florian and {Bielby}, Richard M. and {Bolton}, Adam S. and
         {Brice{\~n}o}, C{\'e}sar and {Buckley-Geer}, Elizabeth J. and
         {Butler}, Karen and {Calamida}, Annalisa and {Carlberg}, Raymond G. and
         {Carter}, Paul and {Casas}, Ricard and {Castander}, Francisco J. and
         {Choi}, Yumi and {Comparat}, Johan and {Cukanovaite}, Elena and
         {Delubac}, Timoth{\'e}e and {DeVries}, Kaitlin and {Dey}, Sharmila and
         {Dhungana}, Govinda and {Dickinson}, Mark and {Ding}, Zhejie and
         {Donaldson}, John B. and {Duan}, Yutong and
         {Duckworth}, Christopher J. and {Eftekharzadeh}, Sarah and
         {Eisenstein}, Daniel J. and {Etourneau}, Thomas and
         {Fagrelius}, Parker A. and {Farihi}, Jay and {Fitzpatrick}, Mike and
         {Font-Ribera}, Andreu and {Fulmer}, Leah and {G{\"a}nsicke}, Boris T. and
         {Gaztanaga}, Enrique and {George}, Koshy and {Gerdes}, David W. and
         {Gontcho}, Satya Gontcho A. and {Gorgoni}, Claudio and
         {Green}, Gregory and {Guy}, Julien and {Harmer}, Diane and {Hernand
        ez}, M. and {Honscheid}, Klaus and {Huang}, Lijuan Wendy and
         {James}, David J. and {Jannuzi}, Buell T. and {Jiang}, Linhua and
         {Joyce}, Richard and {Karcher}, Armin and {Karkar}, Sonia and
         {Kehoe}, Robert and {Kneib}, Jean-Paul and {Kueter-Young}, Andrea and
         {Lan}, Ting-Wen and {Lauer}, Tod R. and {Le Guillou}, Laurent and
         {Le Van Suu}, Auguste and {Lee}, Jae Hyeon and {Lesser}, Michael and
         {Perreault Levasseur}, Laurence and {Li}, Ting S. and
         {Mann}, Justin L. and {Marshall}, Robert and
         {Mart{\'\i}nez-V{\'a}zquez}, C.~E. and {Martini}, Paul and
         {du Mas des Bourboux}, H{\'e}lion and {McManus}, Sean and
         {Meier}, Tobias Gabriel and {M{\'e}nard}, Brice and {Metcalfe}, Nigel and
         {Mu{\~n}oz-Guti{\'e}rrez}, Andrea and {Najita}, Joan and
         {Napier}, Kevin and {Narayan}, Gautham and {Newman}, Jeffrey A. and
         {Nie}, Jundan and {Nord}, Brian and {Norman}, Dara J. and
         {Olsen}, Knut A.~G. and {Paat}, Anthony and
         {Palanque-Delabrouille}, Nathalie and {Peng}, Xiyan and
         {Poppett}, Claire L. and {Poremba}, Megan R. and {Prakash}, Abhishek and
         {Rabinowitz}, David and {Raichoor}, Anand and {Rezaie}, Mehdi and
         {Robertson}, A.~N. and {Roe}, Natalie A. and {Ross}, Ashley J. and
         {Ross}, Nicholas P. and {Rudnick}, Gregory and {Safonova}, Sasha and
         {Saha}, Abhijit and {S{\'a}nchez}, F. Javier and {Savary}, Elodie and
         {Schweiker}, Heidi and {Scott}, Adam and {Seo}, Hee-Jong and
         {Shan}, Huanyuan and {Silva}, David R. and {Slepian}, Zachary and
         {Soto}, Christian and {Sprayberry}, David and {Staten}, Ryan and
         {Stillman}, Coley M. and {Stupak}, Robert J. and {Summers}, David L. and
         {Sien Tie}, Suk and {Tirado}, H. and {Vargas-Maga{\~n}a}, Mariana and
         {Vivas}, A. Katherina and {Wechsler}, Risa H. and {Williams}, Doug and
         {Yang}, Jinyi and {Yang}, Qian and {Yapici}, Tolga and
         {Zaritsky}, Dennis and {Zenteno}, A. and {Zhang}, Kai and
         {Zhang}, Tianmeng and {Zhou}, Rongpu and {Zhou}, Zhimin},
        title = "{Overview of the DESI Legacy Imaging Surveys}",
      journal = {\aj},
         year = 2019,
        month = may,
       volume = {157},
       number = {5},
          eid = {168},
        pages = {168},
          doi = {10.3847/1538-3881/ab089d},
archivePrefix = {arXiv},
       eprint = {1804.08657},
 primaryClass = {astro-ph.IM},
       adsurl = {https://ui.adsabs.harvard.edu/abs/2019AJ....157..168D}
}

@ARTICLE{Eliasdottir_2007,
       author = {{El{\'\i}asd{\'o}ttir}, {\'A}rd{\'\i}s and {Limousin}, Marceau and {Richard}, Johan and {Hjorth}, Jens and {Kneib}, Jean-Paul and {Natarajan}, Priya and {Pedersen}, Kristian and {Jullo}, Eric and {Paraficz}, Danuta},
        title = "{Where is the matter in the Merging Cluster Abell 2218?}",
      journal = {arXiv e-prints},
         year = 2007,
        month = oct,
          eid = {arXiv:0710.5636},
        pages = {arXiv:0710.5636},
archivePrefix = {arXiv},
       eprint = {0710.5636},
 primaryClass = {astro-ph},
       adsurl = {https://ui.adsabs.harvard.edu/abs/2007arXiv0710.5636E}
}

@ARTICLE{Jullo_2007,
       author = {{Jullo}, E. and {Kneib}, J. -P. and {Limousin}, M. and
         {El{\'\i}asd{\'o}ttir}, {\'A}. and {Marshall}, P.~J. and {Verdugo}, T.},
        title = "{A Bayesian approach to strong lensing modelling of galaxy clusters}",
      journal = {New Journal of Physics},
         year = 2007,
        month = dec,
       volume = {9},
       number = {12},
        pages = {447},
          doi = {10.1088/1367-2630/9/12/447},
archivePrefix = {arXiv},
       eprint = {0706.0048},
 primaryClass = {astro-ph},
       adsurl = {https://ui.adsabs.harvard.edu/abs/2007NJPh....9..447J}
}

@ARTICLE{Peng_2002,
       author = {{Peng}, Chien Y. and {Ho}, Luis C. and {Impey}, Chris D. and
         {Rix}, Hans-Walter},
        title = "{Detailed Structural Decomposition of Galaxy Images}",
      journal = {\aj},
         year = 2002,
        month = jul,
       volume = {124},
       number = {1},
        pages = {266-293},
          doi = {10.1086/340952},
archivePrefix = {arXiv},
       eprint = {astro-ph/0204182},
 primaryClass = {astro-ph},
       adsurl = {https://ui.adsabs.harvard.edu/abs/2002AJ....124..266P}
}

@ARTICLE{alexdw_delve_2021,
       author = {{Drlica-Wagner}, A. and {Carlin}, J.~L. and {Nidever}, D.~L. and {Ferguson}, P.~S. and {Kuropatkin}, N. and {Adam{\'o}w}, M. and {Cerny}, W. and {Choi}, Y. and {Esteves}, J.~H. and {Mart{\'\i}nez-V{\'a}zquez}, C.~E. and {Mau}, S. and {Miller}, A.~E. and {Mutlu-Pakdil}, B. and {Neilsen}, E.~H. and {Olsen}, K.~A.~G. and {Pace}, A.~B. and {Riley}, A.~H. and {Sakowska}, J.~D. and {Sand}, D.~J. and {Santana-Silva}, L. and {Tollerud}, E.~J. and {Tucker}, D.~L. and {Vivas}, A.~K. and {Zaborowski}, E. and {Zenteno}, A. and {Abbott}, T.~M.~C. and {Allam}, S. and {Bechtol}, K. and {Bell}, C.~P.~M. and {Bell}, E.~F. and {Bilaji}, P. and {Bom}, C.~R. and {Carballo-Bello}, J.~A. and {Crnojevi{\'c}}, D. and {Cioni}, M.-R.~L. and {Diaz-Ocampo}, A. and {de Boer}, T.~J.~L. and {Erkal}, D. and {Gruendl}, R.~A. and {Hernandez-Lang}, D. and {Hughes}, A.~K. and {James}, D.~J. and {Johnson}, L.~C. and {Li}, T.~S. and {Mao}, Y.-Y. and {Mart{\'\i}nez-Delgado}, D. and {Massana}, P. and {McNanna}, M. and {Morgan}, R. and {Nadler}, E.~O. and {No{\"e}l}, N.~E.~D. and {Palmese}, A. and {Peter}, A.~H.~G. and {Rykoff}, E.~S. and {S{\'a}nchez}, J. and {Shipp}, N. and {Simon}, J.~D. and {Smercina}, A. and {Soares-Santos}, M. and {Stringfellow}, G.~S. and {Tavangar}, K. and {van der Marel}, R.~P. and {Walker}, A.~R. and {Wechsler}, R.~H. and {Wu}, J.~F. and {Yanny}, B. and {Fitzpatrick}, M. and {Huang}, L. and {Jacques}, A. and {Nikutta}, R. and {Scott}, A. and {Astro Data Lab}},
        title = "{The DECam Local Volume Exploration Survey: Overview and First Data Release}",
      journal = {\apjs},
         year = 2021,
        month = sep,
       volume = {256},
       number = {1},
          eid = {2},
        pages = {2},
          doi = {10.3847/1538-4365/ac079d},
archivePrefix = {arXiv},
       eprint = {2103.07476},
 primaryClass = {astro-ph.GA},
       adsurl = {https://ui.adsabs.harvard.edu/abs/2021ApJS..256....2D}
}

@ARTICLE{Zhuang_AGEL_2023,
       author = {{Zhuang}, Zhuyun and {Leethochawalit}, Nicha and {Kirby}, Evan N. and {Nightingale}, J.~W. and {Steidel}, Charles C. and {Glazebrook}, Karl and {Barone}, Tania M. and {Skobe}, Hannah and {Sweet}, Sarah M. and {Nanayakkara}, Themiya and {Allen}, Rebecca J. and {Vasan}, Keerthi G.~C. and {Jones}, Tucker and {Kacprzak}, Glenn G. and {Tran}, Kim-Vy H. and {Jacobs}, Colin},
        title = "{A Glimpse of the Stellar Populations and Elemental Abundances of Gravitationally Lensed, Quiescent Galaxies at z {\ensuremath{\gtrsim}} 1 with Keck Deep Spectroscopy}",
      journal = {\apj},
         year = 2023,
        month = may,
       volume = {948},
       number = {2},
          eid = {132},
        pages = {132},
          doi = {10.3847/1538-4357/acc79b},
archivePrefix = {arXiv},
       eprint = {2212.04731},
 primaryClass = {astro-ph.GA},
       adsurl = {https://ui.adsabs.harvard.edu/abs/2023ApJ...948..132Z}
}

@ARTICLE{donevski2023,
       author = {{Donevski}, D. and {Damjanov}, I. and {Nanni}, A. and {Man}, A. and {Giulietti}, M. and {Romano}, M. and {Lapi}, A. and {Narayanan}, D. and {Dav{\'e}}, R. and {Shivaei}, I. and {Sohn}, J. and {Junais} and {Pantoni}, L. and {Li}, Q.},
        title = "{In pursuit of giants. II. Evolution of dusty quiescent galaxies over the last six billion years from the hCOSMOS survey}",
      journal = {\aap},
         year = 2023,
        month = oct,
       volume = {678},
          eid = {A35},
        pages = {A35},
          doi = {10.1051/0004-6361/202346066},
archivePrefix = {arXiv},
       eprint = {2304.05842},
 primaryClass = {astro-ph.GA},
       adsurl = {https://ui.adsabs.harvard.edu/abs/2023A&A...678A..35D}
}

@INPROCEEDINGS{Tody1986,
       author = {{Tody}, Doug},
        title = "{The IRAF Data Reduction and Analysis System}",
    booktitle = {Instrumentation in astronomy VI},
         year = 1986,
       editor = {{Crawford}, David L.},
       series = {Society of Photo-Optical Instrumentation Engineers (SPIE) Conference Series},
       volume = {627},
        month = jan,
        pages = {733},
          doi = {10.1117/12.968154},
       adsurl = {https://ui.adsabs.harvard.edu/abs/1986SPIE..627..733T}
}

@ARTICLE{Garg_2007,
       author = {{Garg}, Arti and {Stubbs}, Christopher W. and {Challis}, Peter and
         {Wood-Vasey}, W. Michael and {Blondin}, St{\'e}phane and
         {Huber}, Mark E. and {Cook}, Kem and {Nikolaev}, Sergei and
         {Rest}, Armin and {Smith}, R. Chris and {Olsen}, Knut and
         {Suntzeff}, Nicholas B. and {Aguilera}, Claudio and
         {Prieto}, Jose Luis and {Becker}, Andrew and {Miceli}, Antonino and
         {Miknaitis}, Gajus and {Clocchiatti}, Alejandro and {Minniti}, Dante and
         {Morelli}, Lorenzo and {Welch}, Douglas L.},
        title = "{Light Curves of Type Ia Supernovae from Near the Time of Explosion}",
      journal = {\aj},
         year = 2007,
        month = feb,
       volume = {133},
       number = {2},
        pages = {403-419},
          doi = {10.1086/510118},
archivePrefix = {arXiv},
       eprint = {astro-ph/0608639},
 primaryClass = {astro-ph},
       adsurl = {https://ui.adsabs.harvard.edu/abs/2007AJ....133..403G}
}

@ARTICLE{Miknaitis_2007,
       author = {{Miknaitis}, G. and {Pignata}, G. and {Rest}, A. and
         {Wood-Vasey}, W.~M. and {Blondin}, S. and {Challis}, P. and
         {Smith}, R.~C. and {Stubbs}, C.~W. and {Suntzeff}, N.~B. and
         {Foley}, R.~J. and {Matheson}, T. and {Tonry}, J.~L. and
         {Aguilera}, C. and {Blackman}, J.~W. and {Becker}, A.~C. and
         {Clocchiatti}, A. and {Covarrubias}, R. and {Davis}, T.~M. and
         {Filippenko}, A.~V. and {Garg}, A. and {Garnavich}, P.~M. and
         {Hicken}, M. and {Jha}, S. and {Krisciunas}, K. and {Kirshner}, R.~P. and
         {Leibundgut}, B. and {Li}, W. and {Miceli}, A. and {Narayan}, G. and
         {Prieto}, J.~L. and {Riess}, A.~G. and {Salvo}, M.~E. and
         {Schmidt}, B.~P. and {Sollerman}, J. and {Spyromilio}, J. and
         {Zenteno}, A.},
        title = "{The ESSENCE Supernova Survey: Survey Optimization, Observations, and Supernova Photometry}",
      journal = {\apj},
         year = 2007,
        month = sep,
       volume = {666},
       number = {2},
        pages = {674-693},
          doi = {10.1086/519986},
archivePrefix = {arXiv},
       eprint = {astro-ph/0701043},
 primaryClass = {astro-ph},
       adsurl = {https://ui.adsabs.harvard.edu/abs/2007ApJ...666..674M}
}

@ARTICLE{Rest_2005,
       author = {{Rest}, A. and {Stubbs}, C. and {Becker}, A.~C. and {Miknaitis}, G.~A. and
         {Miceli}, A. and {Covarrubias}, R. and {Hawley}, S.~L. and
         {Smith}, R.~C. and {Suntzeff}, N.~B. and {Olsen}, K. and
         {Prieto}, J.~L. and {Hiriart}, R. and {Welch}, D.~L. and {Cook}, K.~H. and
         {Nikolaev}, S. and {Huber}, M. and {Prochtor}, G. and
         {Clocchiatti}, A. and {Minniti}, D. and {Garg}, A. and {Challis}, P. and
         {Keller}, S.~C. and {Schmidt}, B.~P.},
        title = "{Testing LMC Microlensing Scenarios: The Discrimination Power of the SuperMACHO Microlensing Survey}",
      journal = {\apj},
         year = 2005,
        month = dec,
       volume = {634},
       number = {2},
        pages = {1103-1115},
          doi = {10.1086/497060},
archivePrefix = {arXiv},
       eprint = {astro-ph/0509240},
 primaryClass = {astro-ph},
       adsurl = {https://ui.adsabs.harvard.edu/abs/2005ApJ...634.1103R}
}

@INPROCEEDINGS{Tody1993,
       author = {{Tody}, Doug},
        title = "{IRAF in the Nineties}",
    booktitle = {Astronomical Data Analysis Software and Systems II},
         year = 1993,
       editor = {{Hanisch}, R.~J. and {Brissenden}, R.~J.~V. and {Barnes}, J.},
       series = {Astronomical Society of the Pacific Conference Series},
       volume = {52},
        month = jan,
        pages = {173},
       adsurl = {https://ui.adsabs.harvard.edu/abs/1993ASPC...52..173T}
}

@INPROCEEDINGS{Persson_2008,
       author = {{Persson}, S.~E. and {Barkhouser}, Robert and {Birk}, Christoph and
         {Hammond}, Randy and {Harding}, Albert and {Koch}, E.~R. and
         {Marshall}, J.~L. and {McCarthy}, Patrick J. and {Murphy}, David and
         {Orndorff}, Joe and {Scharfstein}, Gregg and {Shectman}, Stephen A. and
         {Smee}, Stephen and {Uomoto}, Alan},
        title = "{The FourStar infrared camera}",
    booktitle = {Ground-based and Airborne Instrumentation for Astronomy II},
         year = 2008,
       series = {Society of Photo-Optical Instrumentation Engineers (SPIE) Conference Series},
       volume = {7014},
        month = jul,
          eid = {70142V},
        pages = {70142V},
          doi = {10.1117/12.790015},
archivePrefix = {arXiv},
       eprint = {0807.3777},
 primaryClass = {astro-ph},
       adsurl = {https://ui.adsabs.harvard.edu/abs/2008SPIE.7014E..2VP}
}

@ARTICLE{lorenzon2025,
       author = {{Lorenzon}, G. and {Donevski}, D. and {Lisiecki}, K. and {Lovell}, C. and {Romano}, M. and {Narayanan}, D. and {Dav{\'e}}, R. and {Man}, A. and {Whitaker}, K.~E. and {Nanni}, A. and {Long}, A. and {Lee}, M.~M. and {Junais} and {Ma{\l}ek}, K. and {Rodighiero}, G. and {Li}, Q.},
        title = "{Tracing the evolutionary pathways of dust and cold gas in high-z quiescent galaxies with SIMBA}",
      journal = {\aap},
         year = 2025,
        month = jan,
       volume = {693},
          eid = {A118},
        pages = {A118},
          doi = {10.1051/0004-6361/202450393},
archivePrefix = {arXiv},
       eprint = {2404.10568},
 primaryClass = {astro-ph.GA},
       adsurl = {https://ui.adsabs.harvard.edu/abs/2025A&A...693A.118L}
}

@ARTICLE{woodrum2022,
       author = {{Woodrum}, Charity and {Williams}, Christina C. and {Rieke}, Marcia and {Leja}, Joel and {Johnson}, Benjamin D. and {Bezanson}, Rachel and {Kennicutt}, Robert and {Spilker}, Justin and {Tacchella}, Sandro},
        title = "{Molecular Gas Reservoirs in Massive Quiescent Galaxies at z   0.7 Linked to Late-time Star Formation}",
      journal = {\apj},
         year = 2022,
        month = nov,
       volume = {940},
       number = {1},
          eid = {39},
        pages = {39},
          doi = {10.3847/1538-4357/ac9af7},
archivePrefix = {arXiv},
       eprint = {2210.03832},
 primaryClass = {astro-ph.GA},
       adsurl = {https://ui.adsabs.harvard.edu/abs/2022ApJ...940...39W}
}

@ARTICLE{scholtz2026,
       author = {{Scholtz}, Jan and {D'Eugenio}, Francesco and {Maiolino}, Roberto and {P{\'e}rez-Gonz{\'a}lez}, Pablo G. and {Circosta}, Chiara and {Tacchella}, Sandro and {Williams}, Christina C. and {Alberts}, Stacey and {Arribas}, Santiago and {Baker}, William M. and {Bertola}, Elena and {Bunker}, Andrew J. and {Carniani}, Stefano and {Charlot}, Stephane and {Cresci}, Giovanni and {Jones}, Gareth C. and {Kumari}, Nimisha and {Lamperti}, Isabella and {Looser}, Tobias J. and {Pino}, Bruno Rodr{\'\i}guez Del and {Robertson}, Brant and {Parlanti}, Eleonora and {Perna}, Michele and {{\"U}bler}, Hannah and {Venturi}, Giacomo and {Witstok}, Joris},
        title = "{Measurement of the gas consumption history of a massive quiescent galaxy}",
      journal = {Nature Astronomy},
         year = 2026,
        month = mar,
       volume = {10},
        pages = {431-439},
          doi = {10.1038/s41550-025-02751-z},
archivePrefix = {arXiv},
       eprint = {2405.19401},
 primaryClass = {astro-ph.GA},
       adsurl = {https://ui.adsabs.harvard.edu/abs/2026NatAs..10..431S}
}

@ARTICLE{setton2025,
       author = {{Setton}, David J. and {Spilker}, Justin S. and {Bezanson}, Rachel and {Suess}, Katherine A. and {Greene}, Jenny E. and {Goulding}, Andy D. and {Cenci}, Elia and {D'Onofrio}, Vincenzo R. and {Feldmann}, Robert and {Kriek}, Mariska and {Kumar}, Anika and {Luo}, Yuanze and {Narayanan}, Desika and {Verrico}, Margaret E. and {Zhu}, Pengpei},
        title = "{SQuIGGL{\textrightarrow}E: Buried Star Formation Cannot Explain the Rapidly Fading CO(2─1) Luminosity in Massive, z {\ensuremath{\sim}} 0.7 Post-starburst Galaxies}",
      journal = {\aj},
         year = 2025,
        month = dec,
       volume = {170},
       number = {6},
          eid = {351},
        pages = {351},
          doi = {10.3847/1538-3881/ae1607},
archivePrefix = {arXiv},
       eprint = {2509.00148},
 primaryClass = {astro-ph.GA},
       adsurl = {https://ui.adsabs.harvard.edu/abs/2025AJ....170..351S}
}

@ARTICLE{hayashi2018,
       author = {{Hayashi}, Masao and {Tadaki}, Ken-ichi and {Kodama}, Tadayuki and {Kohno}, Kotaro and {Yamaguchi}, Yuki and {Hatsukade}, Bunyo and {Koyama}, Yusei and {Shimakawa}, Rhythm and {Tamura}, Yoichi and {Suzuki}, Tomoko L.},
        title = "{Molecular Gas Reservoirs in Cluster Galaxies at z = 1.46}",
      journal = {\apj},
         year = 2018,
        month = apr,
       volume = {856},
       number = {2},
          eid = {118},
        pages = {118},
          doi = {10.3847/1538-4357/aab3e7},
archivePrefix = {arXiv},
       eprint = {1803.00298},
 primaryClass = {astro-ph.GA},
       adsurl = {https://ui.adsabs.harvard.edu/abs/2018ApJ...856..118H}
}

@ARTICLE{donofrio2025,
       author = {{D'Onofrio}, Vincenzo R. and {Spilker}, Justin S. and {Bezanson}, Rachel and {Feldmann}, Robert and {Goulding}, Andy D. and {Greene}, Jenny E. and {Kriek}, Mariska and {Luo}, Yuanze and {Narayanan}, Desika and {Setton}, David J. and {Suess}, Katherine A. and {Zhang}, Yunchong and {Zhu}, Pengpei},
        title = "{Quenching through Tidal Gas Removal: Molecular Gas and Star Formation in Tidal Tails of z {\ensuremath{\sim}} 0.7 Post-starburst Galaxies}",
      journal = {\apj},
         year = 2025,
        month = sep,
       volume = {990},
       number = {2},
          eid = {166},
        pages = {166},
          doi = {10.3847/1538-4357/adf5bd},
archivePrefix = {arXiv},
       eprint = {2507.21249},
 primaryClass = {astro-ph.GA},
       adsurl = {https://ui.adsabs.harvard.edu/abs/2025ApJ...990..166D}
}

@ARTICLE{wu2023,
       author = {{Wu}, Po-Feng and {Bezanson}, Rachel and {D'Eugenio}, Francesco and {Gallazzi}, Anna R. and {Greene}, Jenny E. and {Maseda}, Michael V. and {Suess}, Katherine A. and {van der Wel}, Arjen},
        title = "{Stars, Gas, and Star Formation of Distant Post-starburst Galaxies}",
      journal = {\apj},
         year = 2023,
        month = sep,
       volume = {955},
       number = {1},
          eid = {75},
        pages = {75},
          doi = {10.3847/1538-4357/acf0bd},
archivePrefix = {arXiv},
       eprint = {2308.08681},
 primaryClass = {astro-ph.GA},
       adsurl = {https://ui.adsabs.harvard.edu/abs/2023ApJ...955...75W}
}

@ARTICLE{zanella2023,
       author = {{Zanella}, A. and {Valentino}, F. and {Gallazzi}, A. and {Belli}, S. and {Magdis}, G. and {Bolamperti}, A.},
        title = "{The large molecular gas fraction of post-starburst galaxies at z > 1}",
      journal = {\mnras},
         year = 2023,
        month = sep,
       volume = {524},
       number = {1},
        pages = {923-939},
          doi = {10.1093/mnras/stad1821},
archivePrefix = {arXiv},
       eprint = {2306.08120},
 primaryClass = {astro-ph.GA},
       adsurl = {https://ui.adsabs.harvard.edu/abs/2023MNRAS.524..923Z}
}

@ARTICLE{gobat2018,
       author = {{Gobat}, R. and {Daddi}, E. and {Magdis}, G. and {Bournaud}, F. and {Sargent}, M. and {Martig}, M. and {Jin}, S. and {Finoguenov}, A. and {B{\'e}thermin}, M. and {Hwang}, H.~S. and {Renzini}, A. and {Wilson}, G.~W. and {Aretxaga}, I. and {Yun}, M. and {Strazzullo}, V. and {Valentino}, F.},
        title = "{The unexpectedly large dust and gas content of quiescent galaxies at z > 1.4}",
      journal = {Nature Astronomy},
         year = 2018,
        month = jan,
       volume = {2},
        pages = {239-246},
          doi = {10.1038/s41550-017-0352-5},
archivePrefix = {arXiv},
       eprint = {1703.02207},
 primaryClass = {astro-ph.GA},
       adsurl = {https://ui.adsabs.harvard.edu/abs/2018NatAs...2..239G}
}

@ARTICLE{magdis2021,
       author = {{Magdis}, Georgios E. and {Gobat}, Raphael and {Valentino}, Francesco and {Daddi}, Emanuele and {Zanella}, Anita and {Kokorev}, Vasily and {Toft}, Sune and {Jin}, Shuowen and {Whitaker}, Katherine E.},
        title = "{The interstellar medium of quiescent galaxies and its evolution with time}",
      journal = {\aap},
         year = 2021,
        month = mar,
       volume = {647},
          eid = {A33},
        pages = {A33},
          doi = {10.1051/0004-6361/202039280},
archivePrefix = {arXiv},
       eprint = {2101.04700},
 primaryClass = {astro-ph.GA},
       adsurl = {https://ui.adsabs.harvard.edu/abs/2021A&A...647A..33M}
}

@ARTICLE{blanquezsese2023,
       author = {{Bl{\'a}nquez-Ses{\'e}}, D. and {G{\'o}mez-Guijarro}, C. and {Magdis}, G.~E. and {Magnelli}, B. and {Gobat}, R. and {Daddi}, E. and {Franco}, M. and {Whitaker}, K. and {Valentino}, F. and {Adscheid}, S. and {Schinnerer}, E. and {Zanella}, A. and {Xiao}, M. and {Wang}, T. and {Liu}, D. and {Kokorev}, V. and {Elbaz}, D.},
        title = "{The gas mass reservoir of quiescent galaxies at cosmic noon}",
      journal = {\aap},
         year = 2023,
        month = jun,
       volume = {674},
          eid = {A166},
        pages = {A166},
          doi = {10.1051/0004-6361/202345977},
archivePrefix = {arXiv},
       eprint = {2303.12110},
 primaryClass = {astro-ph.GA},
       adsurl = {https://ui.adsabs.harvard.edu/abs/2023A&A...674A.166B}
}

@ARTICLE{caliendo2021,
       author = {{Caliendo}, Joyce N. and {Whitaker}, Katherine E. and {Akhshik}, Mohammad and {Wilson}, Grant and {Williams}, Christina C. and {Spilker}, Justin S. and {Mahler}, Guillaume and {Pope}, Alexandra and {Sharon}, Keren and {Aguilar}, Emmaly and {Bezanson}, Rachel and {Chavez Dagostino}, Miguel and {G{\'o}mez-Ruiz}, Arturo I. and {Monta{\~n}a}, Alfredo and {Toft}, Sune and {Velazquez de la Rosa}, Miguel and {Zeballos}, Milagros},
        title = "{Early Science with the Large Millimeter Telescope: Constraining the Gas Fraction of a Compact Quiescent Galaxy at z = 1.883}",
      journal = {\apjl},
         year = 2021,
        month = mar,
       volume = {910},
       number = {1},
          eid = {L7},
        pages = {L7},
          doi = {10.3847/2041-8213/abe132},
archivePrefix = {arXiv},
       eprint = {2103.03255},
 primaryClass = {astro-ph.GA},
       adsurl = {https://ui.adsabs.harvard.edu/abs/2021ApJ...910L...7C}
}

@ARTICLE{lee2024,
       author = {{Lee}, Minju M. and {Steidel}, Charles C. and {Brammer}, Gabriel and {F{\"o}rster-Schreiber}, Natascha and {Renzini}, Alvio and {Liu}, Daizhong and {Herrera-Camus}, Rodrigo and {Naab}, Thorsten and {Price}, Sedona H. and {{\"U}bler}, Hannah and {Arriagada-Neira}, Sebasti{\'a}n and {Magdis}, Georgios},
        title = "{High dust content of a quiescent galaxy at z   2 revealed by deep ALMA observation}",
      journal = {\mnras},
         year = 2024,
        month = feb,
       volume = {527},
       number = {4},
        pages = {9529-9547},
          doi = {10.1093/mnras/stad3718},
archivePrefix = {arXiv},
       eprint = {2311.00023},
 primaryClass = {astro-ph.GA},
       adsurl = {https://ui.adsabs.harvard.edu/abs/2024MNRAS.527.9529L}
}

@ARTICLE{whitaker2021,
       author = {{Whitaker}, Katherine E. and {Williams}, Christina C. and {Mowla}, Lamiya and {Spilker}, Justin S. and {Toft}, Sune and {Narayanan}, Desika and {Pope}, Alexandra and {Magdis}, Georgios E. and {van Dokkum}, Pieter G. and {Akhshik}, Mohammad and {Bezanson}, Rachel and {Brammer}, Gabriel B. and {Leja}, Joel and {Man}, Allison and {Nelson}, Erica J. and {Richard}, Johan and {Pacifici}, Camilla and {Sharon}, Keren and {Valentino}, Francesco},
        title = "{Quenching of star formation from a lack of inflowing gas to galaxies}",
      journal = {\nat},
         year = 2021,
        month = sep,
       volume = {597},
       number = {7877},
        pages = {485-488},
          doi = {10.1038/s41586-021-03806-7},
archivePrefix = {arXiv},
       eprint = {2109.10384},
 primaryClass = {astro-ph.GA},
       adsurl = {https://ui.adsabs.harvard.edu/abs/2021Natur.597..485W}
}

@ARTICLE{li2019,
       author = {{Li}, Qi and {Narayanan}, Desika and {Dav{\'e}}, Romeel},
        title = "{The dust-to-gas and dust-to-metal ratio in galaxies from z = 0 to 6}",
      journal = {\mnras},
         year = 2019,
        month = nov,
       volume = {490},
       number = {1},
        pages = {1425-1436},
          doi = {10.1093/mnras/stz2684},
archivePrefix = {arXiv},
       eprint = {1906.09277},
 primaryClass = {astro-ph.GA},
       adsurl = {https://ui.adsabs.harvard.edu/abs/2019MNRAS.490.1425L}
}

@ARTICLE{umehata2025,
       author = {{Umehata}, Hideki and {Kubo}, Mariko and {Nakanishi}, Kouichiro},
        title = "{ADF22-WEB: Detection of a Molecular Gas Reservoir in a Massive Quiescent Galaxy Located in a z ≍ 3 Protocluster Core}",
      journal = {\apjl},
         year = 2025,
        month = may,
       volume = {985},
       number = {1},
          eid = {L8},
        pages = {L8},
          doi = {10.3847/2041-8213/add1d4},
archivePrefix = {arXiv},
       eprint = {2502.06538},
 primaryClass = {astro-ph.GA},
       adsurl = {https://ui.adsabs.harvard.edu/abs/2025ApJ...985L...8U}
}

@ARTICLE{lu2025,
       author = {{Lu}, Shiying and {Daddi}, Emanuele and {Maraston}, Claudia and {Dickinson}, Mark and {Haro}, Pablo Arrabal and {Gobat}, Raphael and {Renzini}, Alvio and {Giavalisco}, Mauro and {Bagley}, Micaela B. and {Calabr{\`o}}, Antonello and {Cheng}, Yingjie and {de la Vega}, Alexander and {D'Eugenio}, Chiara and {Elbaz}, David and {Finkelstein}, Steven L. and {G{\'o}mez-Guijarro}, Carlos and {Gu}, Qiusheng and {Hathi}, Nimish P. and {Huertas-Company}, Marc and {Kartaltepe}, Jeyhan S. and {Koekemoer}, Anton M. and {Henry}, Aur{\'e}lien and {Lyu}, Yipeng and {Magnelli}, Benjamin and {Mobasher}, Bahram and {Papovich}, Casey and {Pirzkal}, Nor and {Rich}, R. Michael and {Tacchella}, Sandro and {Yung}, L.~Y. Aaron},
        title = "{Strong spectral features from asymptotic giant branch stars in distant quiescent galaxies}",
      journal = {Nature Astronomy},
         year = 2025,
        month = jan,
       volume = {9},
        pages = {128-140},
          doi = {10.1038/s41550-024-02391-9},
archivePrefix = {arXiv},
       eprint = {2403.07414},
 primaryClass = {astro-ph.GA},
       adsurl = {https://ui.adsabs.harvard.edu/abs/2025NatAs...9..128L}
}

@ARTICLE{whitaker2026,
       author = {{Whitaker}, Katherine E. and {Bezanson}, Rachel},
        title = "{Quenching of Star Formation in Massive Galaxies}",
      journal = {arXiv e-prints},
         year = 2026,
        month = jun,
          eid = {arXiv:2606.12156},
        pages = {arXiv:2606.12156},
          doi = {10.48550/arXiv.2606.12156},
archivePrefix = {arXiv},
       eprint = {2606.12156},
 primaryClass = {astro-ph.GA},
       adsurl = {https://ui.adsabs.harvard.edu/abs/2026arXiv260612156W}
}

@ARTICLE{whitaker2021b,
       author = {{Whitaker}, Katherine E. and {Narayanan}, Desika and {Williams}, Christina C. and {Li}, Qi and {Spilker}, Justin S. and {Dav{\'e}}, Romeel and {Akhshik}, Mohammad and {Akins}, Hollis B. and {Bezanson}, Rachel and {Katz}, Neal and {Leja}, Joel and {Magdis}, Georgios E. and {Mowla}, Lamiya and {Nelson}, Erica J. and {Pope}, Alexandra and {Privon}, George C. and {Toft}, Sune and {Valentino}, Francesco},
        title = "{High Molecular-gas to Dust Mass Ratios Predicted in Most Quiescent Galaxies}",
      journal = {\apjl},
         year = 2021,
        month = dec,
       volume = {922},
       number = {2},
          eid = {L30},
        pages = {L30},
          doi = {10.3847/2041-8213/ac399f},
archivePrefix = {arXiv},
       eprint = {2111.05349},
 primaryClass = {astro-ph.GA},
       adsurl = {https://ui.adsabs.harvard.edu/abs/2021ApJ...922L..30W}
}

@ARTICLE{spilker2025,
       author = {{Spilker}, Justin S. and {Whitaker}, Katherine E. and {Narayanan}, Desika and {Bezanson}, Rachel and {Bodansky}, Sarah and {D'Onofrio}, Vincenzo R. and {Feldmann}, Robert and {Goulding}, Andy D. and {Greene}, Jenny E. and {Kriek}, Mariska and {Luo}, Yuanze and {Setton}, David J. and {Suess}, Katherine A. and {van der Wel}, Arjen and {Verrico}, Margaret E. and {Williams}, Christina C. and {Woodrum}, Charity and {Wu}, Po-Feng},
        title = "{Unusually High Gas-to-dust Ratios Observed in High-redshift Quiescent Galaxies}",
      journal = {\apjl},
         year = 2025,
        month = nov,
       volume = {993},
       number = {2},
          eid = {L40},
        pages = {L40},
          doi = {10.3847/2041-8213/ae14d6},
archivePrefix = {arXiv},
       eprint = {2507.16914},
 primaryClass = {astro-ph.GA},
       adsurl = {https://ui.adsabs.harvard.edu/abs/2025ApJ...993L..40S}
}

@ARTICLE{akhshik2023,
       author = {{Akhshik}, Mohammad and {Whitaker}, Katherine E. and {Leja}, Joel and {Richard}, Johan and {Spilker}, Justin S. and {Song}, Mimi and {Brammer}, Gabriel and {Bezanson}, Rachel and {Ebeling}, Harald and {Gallazzi}, Anna R. and {Mahler}, Guillaume and {Mowla}, Lamiya A. and {Nelson}, Erica J. and {Pacifici}, Camilla and {Sharon}, Keren and {Toft}, Sune and {Williams}, Christina C. and {Wright}, Lillian and {Zabl}, Johannes},
        title = "{REQUIEM-2D: A Diversity of Formation Pathways in a Sample of Spatially Resolved Massive Quiescent Galaxies at z   2}",
      journal = {\apj},
         year = 2023,
        month = feb,
       volume = {943},
       number = {2},
          eid = {179},
        pages = {179},
          doi = {10.3847/1538-4357/aca677},
archivePrefix = {arXiv},
       eprint = {2203.04979},
 primaryClass = {astro-ph.GA},
       adsurl = {https://ui.adsabs.harvard.edu/abs/2023ApJ...943..179A}
}

@ARTICLE{kalita2021,
       author = {{Kalita}, Boris S. and {Daddi}, Emanuele and {D'Eugenio}, Chiara and {Valentino}, Francesco and {Rich}, R. Michael and {G{\'o}mez-Guijarro}, Carlos and {Coogan}, Rosemary T. and {Delvecchio}, Ivan and {Elbaz}, David and {Neill}, James D. and {Puglisi}, Annagrazia and {Strazzullo}, Veronica},
        title = "{An Ancient Massive Quiescent Galaxy Found in a Gas-rich z   3 Group}",
      journal = {\apjl},
         year = 2021,
        month = aug,
       volume = {917},
       number = {2},
          eid = {L17},
        pages = {L17},
          doi = {10.3847/2041-8213/ac16dc},
archivePrefix = {arXiv},
       eprint = {2107.13241},
 primaryClass = {astro-ph.GA},
       adsurl = {https://ui.adsabs.harvard.edu/abs/2021ApJ...917L..17K}
}

@ARTICLE{suzuki2022,
       author = {{Suzuki}, Tomoko L. and {Glazebrook}, Karl and {Schreiber}, Corentin and {Kodama}, Tadayuki and {Kacprzak}, Glenn G. and {Leiton}, Roger and {Nanayakkara}, Themiya and {Oesch}, Pascal A. and {Papovich}, Casey and {Spitler}, Lee and {Straatman}, Caroline M.~S. and {Tran}, Kim-Vy and {Wang}, Tao},
        title = "{Low Star Formation Activity and Low Gas Content of Quiescent Galaxies at z = 3.5-4.0 Constrained with ALMA}",
      journal = {\apj},
         year = 2022,
        month = sep,
       volume = {936},
       number = {1},
          eid = {61},
        pages = {61},
          doi = {10.3847/1538-4357/ac7ce3},
archivePrefix = {arXiv},
       eprint = {2206.14238},
 primaryClass = {astro-ph.GA},
       adsurl = {https://ui.adsabs.harvard.edu/abs/2022ApJ...936...61S}
}

@ARTICLE{chang2026,
       author = {{Chang}, Wenjun and {Wilson}, Gillian and {Forrest}, Ben and {McConachie}, Ian and {Noble}, Allison and {Muzzin}, Adam and {Marchesini}, Danilo and {Cooper}, Michael C. and {Webb}, Tracy and {Canalizo}, Gabriela and {Gomez}, Percy L. and {Zhu}, Yongda and {Edward}, Adit and {Lei}, Han and {Henry}, Aur{\'e}lien and {Urbano Stawinski}, Stephanie M. and {Wisz}, Marie E.},
        title = "{MAGAZ3NE: Dust Deficiency in Ultramassive Quiescent Galaxies at $3<z<4$ with ALMA Observations}",
      journal = {arXiv e-prints},
         year = 2026,
        month = jan,
          eid = {arXiv:2601.22844},
        pages = {arXiv:2601.22844},
          doi = {10.48550/arXiv.2601.22844},
archivePrefix = {arXiv},
       eprint = {2601.22844},
 primaryClass = {astro-ph.GA},
       adsurl = {https://ui.adsabs.harvard.edu/abs/2026arXiv260122844C}
}

@ARTICLE{siegel2025,
       author = {{Siegel}, Jared C. and {Setton}, David J. and {Greene}, Jenny E. and {Suess}, Katherine A. and {Whitaker}, Katherine E. and {Bezanson}, Rachel and {Leja}, Joel and {Furtak}, Lukas J. and {Cutler}, Sam E. and {de Graaff}, Anna and {Feldmann}, Robert and {Khullar}, Gourav and {Labbe}, Ivo and {Marchesini}, Danilo and {Miller}, Tim B. and {Nanayakkara}, Themiya and {Pan}, Richard and {Price}, Sedona H. and {Treiber}, Helena P. and {van Dokkum}, Pieter and {Wang}, Bingjie and {Weaver}, John R.},
        title = "{UNCOVER: Significant Reddening in Cosmic Noon Quiescent Galaxies}",
      journal = {\apj},
         year = 2025,
        month = may,
       volume = {985},
       number = {1},
          eid = {125},
        pages = {125},
          doi = {10.3847/1538-4357/adc7b7},
archivePrefix = {arXiv},
       eprint = {2409.11457},
 primaryClass = {astro-ph.GA},
       adsurl = {https://ui.adsabs.harvard.edu/abs/2025ApJ...985..125S}
}

@ARTICLE{tacconi2018,
       author = {{Tacconi}, L.~J. and {Genzel}, R. and {Saintonge}, A. and {Combes}, F. and {Garc{\'\i}a-Burillo}, S. and {Neri}, R. and {Bolatto}, A. and {Contini}, T. and {F{\"o}rster Schreiber}, N.~M. and {Lilly}, S. and {Lutz}, D. and {Wuyts}, S. and {Accurso}, G. and {Boissier}, J. and {Boone}, F. and {Bouch{\'e}}, N. and {Bournaud}, F. and {Burkert}, A. and {Carollo}, M. and {Cooper}, M. and {Cox}, P. and {Feruglio}, C. and {Freundlich}, J. and {Herrera-Camus}, R. and {Juneau}, S. and {Lippa}, M. and {Naab}, T. and {Renzini}, A. and {Salome}, P. and {Sternberg}, A. and {Tadaki}, K. and {{\"U}bler}, H. and {Walter}, F. and {Weiner}, B. and {Weiss}, A.},
        title = "{PHIBSS: Unified Scaling Relations of Gas Depletion Time and Molecular Gas Fractions}",
      journal = {\apj},
         year = 2018,
        month = feb,
       volume = {853},
       number = {2},
          eid = {179},
        pages = {179},
          doi = {10.3847/1538-4357/aaa4b4},
archivePrefix = {arXiv},
       eprint = {1702.01140},
 primaryClass = {astro-ph.GA},
       adsurl = {https://ui.adsabs.harvard.edu/abs/2018ApJ...853..179T}
}

@ARTICLE{scoville2016,
       author = {{Scoville}, N. and {Sheth}, K. and {Aussel}, H. and {Vanden Bout}, P. and {Capak}, P. and {Bongiorno}, A. and {Casey}, C.~M. and {Murchikova}, L. and {Koda}, J. and {{\'A}lvarez-M{\'a}rquez}, J. and {Lee}, N. and {Laigle}, C. and {McCracken}, H.~J. and {Ilbert}, O. and {Pope}, A. and {Sanders}, D. and {Chu}, J. and {Toft}, S. and {Ivison}, R.~J. and {Manohar}, S.},
        title = "{ISM Masses and the Star formation Law at Z = 1 to 6: ALMA Observations of Dust Continuum in 145 Galaxies in the COSMOS Survey Field}",
      journal = {\apj},
         year = 2016,
        month = apr,
       volume = {820},
       number = {2},
          eid = {83},
        pages = {83},
          doi = {10.3847/0004-637X/820/2/83},
archivePrefix = {arXiv},
       eprint = {1511.05149},
 primaryClass = {astro-ph.GA},
       adsurl = {https://ui.adsabs.harvard.edu/abs/2016ApJ...820...83S}
}

@ARTICLE{dacunha2021,
       author = {{da Cunha}, E. and {Hodge}, J.~A. and {Casey}, C.~M. and {Algera}, H.~S.~B. and {Kaasinen}, M. and {Smail}, I. and {Walter}, F. and {Brandt}, W.~N. and {Dannerbauer}, H. and {Decarli}, R. and {Groves}, B.~A. and {Knudsen}, K.~K. and {Swinbank}, A.~M. and {Weiss}, A. and {van der Werf}, P. and {Zavala}, J.~A.},
        title = "{Measurements of the Dust Properties in z ≃ 1-3 Submillimeter Galaxies with ALMA}",
      journal = {\apj},
         year = 2021,
        month = sep,
       volume = {919},
       number = {1},
          eid = {30},
        pages = {30},
          doi = {10.3847/1538-4357/ac0ae0},
archivePrefix = {arXiv},
       eprint = {2106.08566},
 primaryClass = {astro-ph.GA},
       adsurl = {https://ui.adsabs.harvard.edu/abs/2021ApJ...919...30D}
}

@ARTICLE{dunne2000,
       author = {{Dunne}, Loretta and {Eales}, Stephen and {Edmunds}, Michael and {Ivison}, Rob and {Alexander}, Paul and {Clements}, David L.},
        title = "{The SCUBA Local Universe Galaxy Survey - I. First measurements of the submillimetre luminosity and dust mass functions}",
      journal = {\mnras},
         year = 2000,
        month = jun,
       volume = {315},
       number = {1},
        pages = {115-139},
          doi = {10.1046/j.1365-8711.2000.03386.x},
archivePrefix = {arXiv},
       eprint = {astro-ph/0002234},
 primaryClass = {astro-ph},
       adsurl = {https://ui.adsabs.harvard.edu/abs/2000MNRAS.315..115D}
}

@ARTICLE{donevski2026,
       author = {{Donevski}, D. and {Nanni}, A. and {Whitaker}, K.~E. and {Man}, A.~W.~S. and {Faisst}, A. and {Lapi}, A. and {Garc{\'\i}a-Bernete}, I. and {Romano}, M. and {Petrushevska}, T. and {Gururajan}, G. and {Lorenzon}, G. and {Narayanan}, D.},
        title = "{Dust and PAHs in late-stage galaxy evolution: Imprints of TP-AGB dust injection, grain growth and AGN feedback in high-z quiescent galaxies with JWST and ALMA}",
      journal = {arXiv e-prints},
         year = 2026,
        month = aug,
          eid = {arXiv:2608.27571},
        pages = {arXiv:2608.27571},
          doi = {10.48550/arXiv.2608.27571},
archivePrefix = {arXiv},
       eprint = {2608.27571},
 primaryClass = {astro-ph.GA},
       adsurl = {https://ui.adsabs.harvard.edu/abs/2026arXiv260827571D}
}

@ARTICLE{greve2012,
       author = {{Greve}, T.~R. and {Vieira}, J.~D. and {Wei{\ss}}, A. and {Aguirre}, J.~E. and {Aird}, K.~A. and {Ashby}, M.~L.~N. and {Benson}, B.~A. and {Bleem}, L.~E. and {Bradford}, C.~M. and {Brodwin}, M. and {Carlstrom}, J.~E. and {Chang}, C.~L. and {Chapman}, S.~C. and {Crawford}, T.~M. and {de Breuck}, C. and {de Haan}, T. and {Dobbs}, M.~A. and {Downes}, T. and {Fassnacht}, C.~D. and {Fazio}, G. and {George}, E.~M. and {Gladders}, M. and {Gonzalez}, A.~H. and {Halverson}, N.~W. and {Hezaveh}, Y. and {High}, F.~W. and {Holder}, G.~P. and {Holzapfel}, W.~L. and {Hoover}, S. and {Hrubes}, J.~D. and {Johnson}, M. and {Keisler}, R. and {Knox}, L. and {Lee}, A.~T. and {Leitch}, E.~M. and {Lueker}, M. and {Luong-Van}, D. and {Malkan}, M. and {Marrone}, D.~P. and {McIntyre}, V. and {McMahon}, J.~J. and {Mehl}, J. and {Menten}, K.~M. and {Meyer}, S.~S. and {Montroy}, T. and {Murphy}, E.~J. and {Natoli}, T. and {Padin}, S. and {Plagge}, T. and {Pryke}, C. and {Reichardt}, C.~L. and {Rest}, A. and {Rosenman}, M. and {Ruel}, J. and {Ruhl}, J.~E. and {Schaffer}, K.~K. and {Sharon}, K. and {Shaw}, L. and {Shirokoff}, E. and {Stalder}, B. and {Stanford}, S.~A. and {Staniszewski}, Z. and {Stark}, A.~A. and {Story}, K. and {Vanderlinde}, K. and {Walsh}, W. and {Welikala}, N. and {Williamson}, R.},
        title = "{Submillimeter Observations of Millimeter Bright Galaxies Discovered by the South Pole Telescope}",
      journal = {\apj},
         year = 2012,
        month = sep,
       volume = {756},
       number = {1},
          eid = {101},
        pages = {101},
          doi = {10.1088/0004-637X/756/1/101},
archivePrefix = {arXiv},
       eprint = {1206.4550},
 primaryClass = {astro-ph.CO},
       adsurl = {https://ui.adsabs.harvard.edu/abs/2012ApJ...756..101G}
}

@ARTICLE{leja2022,
       author = {{Leja}, Joel and {Speagle}, Joshua S. and {Ting}, Yuan-Sen and {Johnson}, Benjamin D. and {Conroy}, Charlie and {Whitaker}, Katherine E. and {Nelson}, Erica J. and {van Dokkum}, Pieter and {Franx}, Marijn},
        title = "{A New Census of the 0.2 < z < 3.0 Universe. II. The Star-forming Sequence}",
      journal = {\apj},
         year = 2022,
        month = sep,
       volume = {936},
       number = {2},
          eid = {165},
        pages = {165},
          doi = {10.3847/1538-4357/ac887d},
archivePrefix = {arXiv},
       eprint = {2110.04314},
 primaryClass = {astro-ph.GA},
       adsurl = {https://ui.adsabs.harvard.edu/abs/2022ApJ...936..165L}
}

@ARTICLE{michalowski2010,
       author = {{Micha{\l}owski}, Micha{\l} J. and {Watson}, Darach and {Hjorth}, Jens},
        title = "{Rapid Dust Production in Submillimeter Galaxies at z > 4?}",
      journal = {\apj},
         year = 2010,
        month = apr,
       volume = {712},
       number = {2},
        pages = {942-950},
          doi = {10.1088/0004-637X/712/2/942},
archivePrefix = {arXiv},
       eprint = {1002.2636},
 primaryClass = {astro-ph.CO},
       adsurl = {https://ui.adsabs.harvard.edu/abs/2010ApJ...712..942M}
}

@ARTICLE{salpeter1955,
       author = {{Salpeter}, Edwin E.},
        title = "{The Luminosity Function and Stellar Evolution}",
      journal = {\apj},
         year = 1955,
        month = jan,
       volume = {121},
        pages = {161},
          doi = {10.1086/145971},
       adsurl = {https://ui.adsabs.harvard.edu/abs/1955ApJ...121..161S}
}

@ARTICLE{davies2024,
       author = {{Davies}, Rebecca L. and {Belli}, Sirio and {Park}, Minjung and {Mendel}, J. Trevor and {Johnson}, Benjamin D. and {Conroy}, Charlie and {Benton}, Chloe and {Bugiani}, Letizia and {Emami}, Razieh and {Leja}, Joel and {Li}, Yijia and {Maheson}, Gabriel and {Mathews}, Elijah P. and {Naidu}, Rohan P. and {Nelson}, Erica J. and {Tacchella}, Sandro and {Terrazas}, Bryan A. and {Weinberger}, Rainer},
        title = "{JWST reveals widespread AGN driven neutral gas outflows in massive z2 galaxies}",
      journal = {\mnras},
         year = 2024,
        month = mar,
       volume = {528},
       number = {3},
        pages = {4976-4992},
          doi = {10.1093/mnras/stae327},
archivePrefix = {arXiv},
       eprint = {2310.17939},
 primaryClass = {astro-ph.GA},
       adsurl = {https://ui.adsabs.harvard.edu/abs/2024MNRAS.528.4976D}
}

@ARTICLE{belli2024,
       author = {{Belli}, Sirio and {Park}, Minjung and {Davies}, Rebecca L. and {Mendel}, J. Trevor and {Johnson}, Benjamin D. and {Conroy}, Charlie and {Benton}, Chloe and {Bugiani}, Letizia and {Emami}, Razieh and {Leja}, Joel and {Li}, Yijia and {Maheson}, Gabriel and {Mathews}, Elijah P. and {Naidu}, Rohan P. and {Nelson}, Erica J. and {Tacchella}, Sandro and {Terrazas}, Bryan A. and {Weinberger}, Rainer},
        title = "{Star formation shut down by multiphase gas outflow in a galaxy at a redshift of 2.45}",
      journal = {\nat},
         year = 2024,
        month = jun,
       volume = {630},
       number = {8015},
        pages = {54-58},
          doi = {10.1038/s41586-024-07412-1},
archivePrefix = {arXiv},
       eprint = {2308.05795},
 primaryClass = {astro-ph.GA},
       adsurl = {https://ui.adsabs.harvard.edu/abs/2024Natur.630...54B}
}

@ARTICLE{zhu2026,
       author = {{Zhu}, Pengpei and {Ito}, Kei and {Valentino}, Francesco and {Hamadouche}, Massissilia and {Scarpe}, Gianluca and {Whitaker}, Katherine E. and {Kakimoto}, Takumi and {Baker}, William M. and {Gallazzi}, Anna R. and {Gillman}, Steven and {Gottumukkala}, Rashmi and {Jespersen}, Christian Kragh and {Lee}, Minju and {Man}, Allison W.~S. and {Magdis}, Georgios and {Onodera}, Masato and {Shimakawa}, Rhythm and {Vijayan}, Aswin and {Wu}, Po-Feng},
        title = "{There and back again: Neutral outflows in z \raisebox{-0.5ex}\textasciitilde 3.5 quiescent galaxies}",
      journal = {\aap},
         year = 2026,
        month = aug,
       volume = {712},
          eid = {A169},
        pages = {A169},
          doi = {10.1051/0004-6361/202659510},
archivePrefix = {arXiv},
       eprint = {2602.17767},
 primaryClass = {astro-ph.GA},
       adsurl = {https://ui.adsabs.harvard.edu/abs/2026A&A...712A.169Z}
}

@ARTICLE{bugiani2025,
       author = {{Bugiani}, Letizia and {Belli}, Sirio and {Park}, Minjung and {Davies}, Rebecca L. and {Mendel}, J. Trevor and {Johnson}, Benjamin D. and {Khoram}, Amir H. and {Benton}, Chloe and {Cimatti}, Andrea and {Conroy}, Charlie and {Emami}, Razieh and {Leja}, Joel and {Li}, Yijia and {Maheson}, Gabriel and {Mathews}, Elijah P. and {Naidu}, Rohan P. and {Nelson}, Erica J. and {Tacchella}, Sandro and {Terrazas}, Bryan A. and {Weinberger}, Rainer},
        title = "{Active Galactic Nucleus Feedback in Quiescent Galaxies at Cosmic Noon Traced by Ionized Gas Emission}",
      journal = {\apj},
         year = 2025,
        month = mar,
       volume = {981},
       number = {1},
          eid = {25},
        pages = {25},
          doi = {10.3847/1538-4357/adaeaf},
archivePrefix = {arXiv},
       eprint = {2406.08547},
 primaryClass = {astro-ph.GA},
       adsurl = {https://ui.adsabs.harvard.edu/abs/2025ApJ...981...25B}
}

@ARTICLE{salim2020,
       author = {{Salim}, Samir and {Narayanan}, Desika},
        title = "{The Dust Attenuation Law in Galaxies}",
      journal = {\araa},
         year = 2020,
        month = aug,
       volume = {58},
        pages = {529-575},
          doi = {10.1146/annurev-astro-032620-021933},
archivePrefix = {arXiv},
       eprint = {2001.03181},
 primaryClass = {astro-ph.GA},
       adsurl = {https://ui.adsabs.harvard.edu/abs/2020ARA&A..58..529S}
}

@ARTICLE{muzzin2013,
       author = {{Muzzin}, Adam and {Marchesini}, Danilo and {Stefanon}, Mauro and {Franx}, Marijn and {Milvang-Jensen}, Bo and {Dunlop}, James S. and {Fynbo}, J.~P.~U. and {Brammer}, Gabriel and {Labb{\'e}}, Ivo and {van Dokkum}, Pieter},
        title = "{A Public K$_{s}$ -selected Catalog in the COSMOS/ULTRAVISTA Field: Photometry, Photometric Redshifts, and Stellar Population Parameters}",
      journal = {\apjs},
         year = 2013,
        month = may,
       volume = {206},
       number = {1},
          eid = {8},
        pages = {8},
          doi = {10.1088/0067-0049/206/1/8},
archivePrefix = {arXiv},
       eprint = {1303.4410},
 primaryClass = {astro-ph.CO},
       adsurl = {https://ui.adsabs.harvard.edu/abs/2013ApJS..206....8M}
}

@ARTICLE{ilbert2013,
       author = {{Ilbert}, O. and {McCracken}, H.~J. and {Le F{\`e}vre}, O. and {Capak}, P. and {Dunlop}, J. and {Karim}, A. and {Renzini}, M.~A. and {Caputi}, K. and {Boissier}, S. and {Arnouts}, S. and {Aussel}, H. and {Comparat}, J. and {Guo}, Q. and {Hudelot}, P. and {Kartaltepe}, J. and {Kneib}, J.~P. and {Krogager}, J.~K. and {Le Floc'h}, E. and {Lilly}, S. and {Mellier}, Y. and {Milvang-Jensen}, B. and {Moutard}, T. and {Onodera}, M. and {Richard}, J. and {Salvato}, M. and {Sanders}, D.~B. and {Scoville}, N. and {Silverman}, J.~D. and {Taniguchi}, Y. and {Tasca}, L. and {Thomas}, R. and {Toft}, S. and {Tresse}, L. and {Vergani}, D. and {Wolk}, M. and {Zirm}, A.},
        title = "{Mass assembly in quiescent and star-forming galaxies since z ≃ 4 from UltraVISTA}",
      journal = {\aap},
         year = 2013,
        month = aug,
       volume = {556},
          eid = {A55},
        pages = {A55},
          doi = {10.1051/0004-6361/201321100},
archivePrefix = {arXiv},
       eprint = {1301.3157},
 primaryClass = {astro-ph.CO},
       adsurl = {https://ui.adsabs.harvard.edu/abs/2013A&A...556A..55I}
}

@ARTICLE{Tomczak2014,
       author = {{Tomczak}, Adam R. and {Quadri}, Ryan F. and {Tran}, Kim-Vy H. and {Labb{\'e}}, Ivo and {Straatman}, Caroline M.~S. and {Papovich}, Casey and {Glazebrook}, Karl and {Allen}, Rebecca and {Brammer}, Gabriel B. and {Kacprzak}, Glenn G. and {Kawinwanichakij}, Lalitwadee and {Kelson}, Daniel D. and {McCarthy}, Patrick J. and {Mehrtens}, Nicola and {Monson}, Andrew J. and {Persson}, S. Eric and {Spitler}, Lee R. and {Tilvi}, Vithal and {van Dokkum}, Pieter},
        title = "{Galaxy Stellar Mass Functions from ZFOURGE/CANDELS: An Excess of Low-mass Galaxies since z = 2 and the Rapid Buildup of Quiescent Galaxies}",
      journal = {\apj},
         year = 2014,
        month = mar,
       volume = {783},
       number = {2},
          eid = {85},
        pages = {85},
          doi = {10.1088/0004-637X/783/2/85},
archivePrefix = {arXiv},
       eprint = {1309.5972},
 primaryClass = {astro-ph.CO},
       adsurl = {https://ui.adsabs.harvard.edu/abs/2014ApJ...783...85T}
}

@ARTICLE{muzzin2013a,
       author = {{Muzzin}, Adam and {Marchesini}, Danilo and {Stefanon}, Mauro and {Franx}, Marijn and {McCracken}, Henry J. and {Milvang-Jensen}, Bo and {Dunlop}, James S. and {Fynbo}, J.~P.~U. and {Brammer}, Gabriel and {Labb{\'e}}, Ivo and {van Dokkum}, Pieter G.},
        title = "{The Evolution of the Stellar Mass Functions of Star-forming and Quiescent Galaxies to z = 4 from the COSMOS/UltraVISTA Survey}",
      journal = {\apj},
         year = 2013,
        month = nov,
       volume = {777},
       number = {1},
          eid = {18},
        pages = {18},
          doi = {10.1088/0004-637X/777/1/18},
archivePrefix = {arXiv},
       eprint = {1303.4409},
 primaryClass = {astro-ph.CO},
       adsurl = {https://ui.adsabs.harvard.edu/abs/2013ApJ...777...18M}
}

@ARTICLE{Davidzon2017,
       author = {{Davidzon}, I. and {Ilbert}, O. and {Laigle}, C. and {Coupon}, J. and {McCracken}, H.~J. and {Delvecchio}, I. and {Masters}, D. and {Capak}, P. and {Hsieh}, B.~C. and {Le F{\`e}vre}, O. and {Tresse}, L. and {Bethermin}, M. and {Chang}, Y.-Y. and {Faisst}, A.~L. and {Le Floc'h}, E. and {Steinhardt}, C. and {Toft}, S. and {Aussel}, H. and {Dubois}, C. and {Hasinger}, G. and {Salvato}, M. and {Sanders}, D.~B. and {Scoville}, N. and {Silverman}, J.~D.},
        title = "{The COSMOS2015 galaxy stellar mass function . Thirteen billion years of stellar mass assembly in ten snapshots}",
      journal = {\aap},
         year = 2017,
        month = sep,
       volume = {605},
          eid = {A70},
        pages = {A70},
          doi = {10.1051/0004-6361/201730419},
archivePrefix = {arXiv},
       eprint = {1701.02734},
 primaryClass = {astro-ph.GA},
       adsurl = {https://ui.adsabs.harvard.edu/abs/2017A&A...605A..70D}
}

@ARTICLE{daddi2005,
       author = {{Daddi}, E. and {Renzini}, A. and {Pirzkal}, N. and {Cimatti}, A. and {Malhotra}, S. and {Stiavelli}, M. and {Xu}, C. and {Pasquali}, A. and {Rhoads}, J.~E. and {Brusa}, M. and {di Serego Alighieri}, S. and {Ferguson}, H.~C. and {Koekemoer}, A.~M. and {Moustakas}, L.~A. and {Panagia}, N. and {Windhorst}, R.~A.},
        title = "{Passively Evolving Early-Type Galaxies at 1.4 <\raisebox{-0.5ex}\textasciitilde z <\raisebox{-0.5ex}\textasciitilde 2.5 in the Hubble Ultra Deep Field}",
      journal = {\apj},
         year = 2005,
        month = jun,
       volume = {626},
       number = {2},
        pages = {680-697},
          doi = {10.1086/430104},
archivePrefix = {arXiv},
       eprint = {astro-ph/0503102},
 primaryClass = {astro-ph},
       adsurl = {https://ui.adsabs.harvard.edu/abs/2005ApJ...626..680D}
}

@ARTICLE{vandokkum2008,
       author = {{van Dokkum}, Pieter G. and {Franx}, Marijn and {Kriek}, Mariska and {Holden}, Bradford and {Illingworth}, Garth D. and {Magee}, Daniel and {Bouwens}, Rychard and {Marchesini}, Danilo and {Quadri}, Ryan and {Rudnick}, Greg and {Taylor}, Edward N. and {Toft}, Sune},
        title = "{Confirmation of the Remarkable Compactness of Massive Quiescent Galaxies at z \raisebox{-0.5ex}\textasciitilde 2.3: Early-Type Galaxies Did not Form in a Simple Monolithic Collapse}",
      journal = {\apjl},
         year = 2008,
        month = apr,
       volume = {677},
       number = {1},
        pages = {L5},
          doi = {10.1086/587874},
archivePrefix = {arXiv},
       eprint = {0802.4094},
 primaryClass = {astro-ph},
       adsurl = {https://ui.adsabs.harvard.edu/abs/2008ApJ...677L...5V}
}

@ARTICLE{vanderwel2014,
       author = {{van der Wel}, A. and {Franx}, M. and {van Dokkum}, P.~G. and {Skelton}, R.~E. and {Momcheva}, I.~G. and {Whitaker}, K.~E. and {Brammer}, G.~B. and {Bell}, E.~F. and {Rix}, H.-W. and {Wuyts}, S. and {Ferguson}, H.~C. and {Holden}, B.~P. and {Barro}, G. and {Koekemoer}, A.~M. and {Chang}, Yu-Yen and {McGrath}, E.~J. and {H{\"a}ussler}, B. and {Dekel}, A. and {Behroozi}, P. and {Fumagalli}, M. and {Leja}, J. and {Lundgren}, B.~F. and {Maseda}, M.~V. and {Nelson}, E.~J. and {Wake}, D.~A. and {Patel}, S.~G. and {Labb{\'e}}, I. and {Faber}, S.~M. and {Grogin}, N.~A. and {Kocevski}, D.~D.},
        title = "{3D-HST+CANDELS: The Evolution of the Galaxy Size-Mass Distribution since z = 3}",
      journal = {\apj},
         year = 2014,
        month = jun,
       volume = {788},
       number = {1},
          eid = {28},
        pages = {28},
          doi = {10.1088/0004-637X/788/1/28},
archivePrefix = {arXiv},
       eprint = {1404.2844},
 primaryClass = {astro-ph.GA},
       adsurl = {https://ui.adsabs.harvard.edu/abs/2014ApJ...788...28V}
}

@ARTICLE{kriek2006,
       author = {{Kriek}, Mariska and {van Dokkum}, Pieter G. and {Franx}, Marijn and {F{\"o}rster Schreiber}, Natascha M. and {Gawiser}, Eric and {Illingworth}, Garth D. and {Labb{\'e}}, Ivo and {Marchesini}, Danilo and {Quadri}, Ryan and {Rix}, Hans-Walter and {Rudnick}, Gregory and {Toft}, Sune and {van der Werf}, Paul and {Wuyts}, Stijn},
        title = "{Direct Measurements of the Stellar Continua and Balmer/4000 {\r{A}} Breaks of Red z > 2 Galaxies: Redshifts and Improved Constraints on Stellar Populations}",
      journal = {\apj},
         year = 2006,
        month = jul,
       volume = {645},
       number = {1},
        pages = {44-54},
          doi = {10.1086/504103},
archivePrefix = {arXiv},
       eprint = {astro-ph/0603063},
 primaryClass = {astro-ph},
       adsurl = {https://ui.adsabs.harvard.edu/abs/2006ApJ...645...44K}
}

@ARTICLE{williams2009,
       author = {{Williams}, Rik J. and {Quadri}, Ryan F. and {Franx}, Marijn and {van Dokkum}, Pieter and {Labb{\'e}}, Ivo},
        title = "{Detection of Quiescent Galaxies in a Bicolor Sequence from Z = 0-2}",
      journal = {\apj},
         year = 2009,
        month = feb,
       volume = {691},
       number = {2},
        pages = {1879-1895},
          doi = {10.1088/0004-637X/691/2/1879},
archivePrefix = {arXiv},
       eprint = {0806.0625},
 primaryClass = {astro-ph},
       adsurl = {https://ui.adsabs.harvard.edu/abs/2009ApJ...691.1879W}
}

@ARTICLE{whitaker2011,
       author = {{Whitaker}, Katherine E. and {Labb{\'e}}, Ivo and {van Dokkum}, Pieter G. and {Brammer}, Gabriel and {Kriek}, Mariska and {Marchesini}, Danilo and {Quadri}, Ryan F. and {Franx}, Marijn and {Muzzin}, Adam and {Williams}, Rik J. and {Bezanson}, Rachel and {Illingworth}, Garth D. and {Lee}, Kyoung-Soo and {Lundgren}, Britt and {Nelson}, Erica J. and {Rudnick}, Gregory and {Tal}, Tomer and {Wake}, David A.},
        title = "{The NEWFIRM Medium-band Survey: Photometric Catalogs, Redshifts, and the Bimodal Color Distribution of Galaxies out to z \raisebox{-0.5ex}\textasciitilde 3}",
      journal = {\apj},
         year = 2011,
        month = jul,
       volume = {735},
       number = {2},
          eid = {86},
        pages = {86},
          doi = {10.1088/0004-637X/735/2/86},
archivePrefix = {arXiv},
       eprint = {1105.4609},
 primaryClass = {astro-ph.CO},
       adsurl = {https://ui.adsabs.harvard.edu/abs/2011ApJ...735...86W}
}

@ARTICLE{whitaker2013,
       author = {{Whitaker}, Katherine E. and {van Dokkum}, Pieter G. and {Brammer}, Gabriel and {Momcheva}, Ivelina G. and {Skelton}, Rosalind and {Franx}, Marijn and {Kriek}, Mariska and {Labb{\'e}}, Ivo and {Fumagalli}, Mattia and {Lundgren}, Britt F. and {Nelson}, Erica J. and {Patel}, Shannon G. and {Rix}, Hans-Walter},
        title = "{Quiescent Galaxies in the 3D-HST Survey: Spectroscopic Confirmation of a Large Number of Galaxies with Relatively Old Stellar Populations at z \raisebox{-0.5ex}\textasciitilde 2}",
      journal = {\apjl},
         year = 2013,
        month = jun,
       volume = {770},
       number = {2},
          eid = {L39},
        pages = {L39},
          doi = {10.1088/2041-8205/770/2/L39},
archivePrefix = {arXiv},
       eprint = {1305.1943},
 primaryClass = {astro-ph.CO},
       adsurl = {https://ui.adsabs.harvard.edu/abs/2013ApJ...770L..39W}
}

@ARTICLE{bluck2026a,
       author = {{Bluck}, Asa F.~L.},
        title = "{A Review of Galaxy Quenching -- Part I: Defining the Problem and Observational Results}",
      journal = {arXiv e-prints},
         year = 2026,
        month = aug,
          eid = {arXiv:2608.07221},
        pages = {arXiv:2608.07221},
          doi = {10.48550/arXiv.2608.07221},
archivePrefix = {arXiv},
       eprint = {2608.07221},
 primaryClass = {astro-ph.GA},
       adsurl = {https://ui.adsabs.harvard.edu/abs/2026arXiv260807221B}
}

@ARTICLE{bluck2026b,
       author = {{Bluck}, Asa F.~L.},
        title = "{A Review of Galaxy Quenching -- Part II: Theoretical Solutions and Direct Observational Tests}",
      journal = {arXiv e-prints},
         year = 2026,
        month = aug,
          eid = {arXiv:2608.07247},
        pages = {arXiv:2608.07247},
          doi = {10.48550/arXiv.2608.07247},
archivePrefix = {arXiv},
       eprint = {2608.07247},
 primaryClass = {astro-ph.GA},
       adsurl = {https://ui.adsabs.harvard.edu/abs/2026arXiv260807247B}
}

@ARTICLE{gomez2010,
       author = {{Gomez}, H.~L. and {Baes}, M. and {Cortese}, L. and {Smith}, M.~W.~L. and {Boselli}, A. and {Ciesla}, L. and {Bendo}, G.~J. and {Pohlen}, M. and {di Serego Alighieri}, S. and {Auld}, R. and {Barlow}, M.~J. and {Bock}, J.~J. and {Bradford}, M. and {Buat}, V. and {Castro-Rodriguez}, N. and {Chanial}, P. and {Charlot}, S. and {Clements}, D.~L. and {Cooray}, A. and {Cormier}, D. and {Davies}, J.~I. and {Dwek}, E. and {Eales}, S. and {Elbaz}, D. and {Galametz}, M. and {Galliano}, F. and {Gear}, W.~K. and {Glenn}, J. and {Griffin}, M. and {Hony}, S. and {Isaak}, K.~G. and {Levenson}, L.~R. and {Lu}, N. and {Madden}, S. and {O'Halloran}, B. and {Okumura}, K. and {Oliver}, S. and {Page}, M.~J. and {Panuzzo}, P. and {Papageorgiou}, A. and {Parkin}, T.~J. and {Perez-Fournon}, I. and {Rangwala}, N. and {Rigby}, E.~E. and {Roussel}, H. and {Rykala}, A. and {Sacchi}, N. and {Sauvage}, M. and {Schirm}, M.~R.~P. and {Schulz}, B. and {Spinoglio}, L. and {Srinivasan}, S. and {Stevens}, J.~A. and {Symeonidis}, M. and {Trichas}, M. and {Vaccari}, M. and {Vigroux}, L. and {Wilson}, C.~D. and {Wozniak}, H. and {Wright}, G.~S. and {Zeilinger}, W.~W.},
        title = "{The dust morphology of the elliptical Galaxy M 86 with SPIRE}",
      journal = {\aap},
         year = 2010,
        month = jul,
       volume = {518},
          eid = {L45},
        pages = {L45},
          doi = {10.1051/0004-6361/201014530},
archivePrefix = {arXiv},
       eprint = {1005.1597},
 primaryClass = {astro-ph.CO},
       adsurl = {https://ui.adsabs.harvard.edu/abs/2010A&A...518L..45G}
}

@ARTICLE{rowlands2012,
       author = {{Rowlands}, K. and {Dunne}, L. and {Maddox}, S. and {Bourne}, N. and {Gomez}, H.~L. and {Kaviraj}, S. and {Bamford}, S.~P. and {Brough}, S. and {Charlot}, S. and {da Cunha}, E. and {Driver}, S.~P. and {Eales}, S.~A. and {Hopkins}, A.~M. and {Kelvin}, L. and {Nichol}, R.~C. and {Sansom}, A.~E. and {Sharp}, R. and {Smith}, D.~J.~B. and {Temi}, P. and {van der Werf}, P. and {Baes}, M. and {Cava}, A. and {Cooray}, A. and {Croom}, S.~M. and {Dariush}, A. and {de Zotti}, G. and {Dye}, S. and {Fritz}, J. and {Hopwood}, R. and {Ibar}, E. and {Ivison}, R.~J. and {Liske}, J. and {Loveday}, J. and {Madore}, B. and {Norberg}, P. and {Popescu}, C.~C. and {Rigby}, E.~E. and {Robotham}, A. and {Rodighiero}, G. and {Seibert}, M. and {Tuffs}, R.~J.},
        title = "{Herschel-ATLAS/GAMA: dusty early-type galaxies and passive spirals}",
      journal = {\mnras},
         year = 2012,
        month = jan,
       volume = {419},
       number = {3},
        pages = {2545-2578},
          doi = {10.1111/j.1365-2966.2011.19905.x},
archivePrefix = {arXiv},
       eprint = {1109.6274},
 primaryClass = {astro-ph.CO},
       adsurl = {https://ui.adsabs.harvard.edu/abs/2012MNRAS.419.2545R}
}

@ARTICLE{Lianou2016,
       author = {{Lianou}, S. and {Xilouris}, E. and {Madden}, S.~C. and {Barmby}, P.},
        title = "{The dustier early-type galaxies deviate from late-type galaxies' scaling relations}",
      journal = {\mnras},
         year = 2016,
        month = sep,
       volume = {461},
       number = {3},
        pages = {2856-2866},
          doi = {10.1093/mnras/stw1467},
archivePrefix = {arXiv},
       eprint = {1606.04867},
 primaryClass = {astro-ph.GA},
       adsurl = {https://ui.adsabs.harvard.edu/abs/2016MNRAS.461.2856L}
}

@ARTICLE{michalowski2019,
       author = {{Micha{\l}owski}, M.~J. and {Hjorth}, J. and {Gall}, C. and {Frayer}, D.~T. and {Tsai}, A.-L. and {Hirashita}, H. and {Rowlands}, K. and {Takeuchi}, T.~T. and {Le{\'s}niewska}, A. and {Behrendt}, D. and {Bourne}, N. and {Hughes}, D.~H. and {Spring}, E. and {Zavala}, J. and {Bartczak}, P.},
        title = "{The fate of the interstellar medium in early-type galaxies. I. First direct measurement of the timescale of dust removal}",
      journal = {\aap},
         year = 2019,
        month = dec,
       volume = {632},
          eid = {A43},
        pages = {A43},
          doi = {10.1051/0004-6361/201936055},
archivePrefix = {arXiv},
       eprint = {1910.06327},
 primaryClass = {astro-ph.GA},
       adsurl = {https://ui.adsabs.harvard.edu/abs/2019A&A...632A..43M}
}

@ARTICLE{lesniewska2023,
       author = {{Le{\'s}niewska}, Aleksandra and {Micha{\l}owski}, M.~J. and {Gall}, C. and {Hjorth}, J. and {Nadolny}, J. and {Ryzhov}, O. and {Solar}, M.},
        title = "{The Fate of the Interstellar Medium in Early-type Galaxies. II. Observational Evidence for Morphological Quenching}",
      journal = {\apj},
         year = 2023,
        month = aug,
       volume = {953},
       number = {1},
          eid = {27},
        pages = {27},
          doi = {10.3847/1538-4357/acdcfc},
archivePrefix = {arXiv},
       eprint = {2306.05774},
 primaryClass = {astro-ph.GA},
       adsurl = {https://ui.adsabs.harvard.edu/abs/2023ApJ...953...27L}
}

@ARTICLE{michalowski2024,
       author = {{Micha{\l}owski}, Micha{\l} J. and {Gall}, C. and {Hjorth}, J. and {Frayer}, D.~T. and {Tsai}, A.-L. and {Rowlands}, K. and {Takeuchi}, T.~T. and {Le{\'s}niewska}, A. and {Behrendt}, D. and {Bourne}, N. and {Hughes}, D.~H. and {Koprowski}, M.~P. and {Nadolny}, J. and {Ryzhov}, O. and {Solar}, M. and {Spring}, E. and {Zavala}, J. and {Bartczak}, P.},
        title = "{The Fate of the Interstellar Medium in Early-type Galaxies. III. The Mechanism of Interstellar Medium Removal and the Quenching of Star Formation}",
      journal = {\apj},
         year = 2024,
        month = apr,
       volume = {964},
       number = {2},
          eid = {129},
        pages = {129},
          doi = {10.3847/1538-4357/ad1b52},
archivePrefix = {arXiv},
       eprint = {2401.04774},
 primaryClass = {astro-ph.GA},
       adsurl = {https://ui.adsabs.harvard.edu/abs/2024ApJ...964..129M}
}

@ARTICLE{scoville2014,
       author = {{Scoville}, N. and {Aussel}, H. and {Sheth}, K. and {Scott}, K.~S. and {Sanders}, D. and {Ivison}, R. and {Pope}, A. and {Capak}, P. and {Vanden Bout}, P. and {Manohar}, S. and {Kartaltepe}, J. and {Robertson}, B. and {Lilly}, S.},
        title = "{The Evolution of Interstellar Medium Mass Probed by Dust Emission: ALMA Observations at z = 0.3-2}",
      journal = {\apj},
         year = 2014,
        month = mar,
       volume = {783},
       number = {2},
          eid = {84},
        pages = {84},
          doi = {10.1088/0004-637X/783/2/84},
archivePrefix = {arXiv},
       eprint = {1401.2987},
 primaryClass = {astro-ph.GA},
       adsurl = {https://ui.adsabs.harvard.edu/abs/2014ApJ...783...84S}
}

@ARTICLE{tacchella2015,
       author = {{Tacchella}, S. and {Carollo}, C.~M. and {Renzini}, A. and {F{\"o}rster Schreiber}, N.~M. and {Lang}, P. and {Wuyts}, S. and {Cresci}, G. and {Dekel}, A. and {Genzel}, R. and {Lilly}, S.~J. and {Mancini}, C. and {Newman}, S. and {Onodera}, M. and {Shapley}, A. and {Tacconi}, L. and {Woo}, J. and {Zamorani}, G.},
        title = "{Evidence for mature bulges and an inside-out quenching phase 3 billion years after the Big Bang}",
      journal = {Science},
         year = 2015,
        month = apr,
       volume = {348},
       number = {6232},
        pages = {314-317},
          doi = {10.1126/science.1261094},
archivePrefix = {arXiv},
       eprint = {1504.04021},
 primaryClass = {astro-ph.GA},
       adsurl = {https://ui.adsabs.harvard.edu/abs/2015Sci...348..314T}
}

@ARTICLE{tacchella2018,
       author = {{Tacchella}, S. and {Carollo}, C.~M. and {F{\"o}rster Schreiber}, N.~M. and {Renzini}, A. and {Dekel}, A. and {Genzel}, R. and {Lang}, P. and {Lilly}, S.~J. and {Mancini}, C. and {Onodera}, M. and {Tacconi}, L.~J. and {Wuyts}, S. and {Zamorani}, G.},
        title = "{Dust Attenuation, Bulge Formation, and Inside-out Quenching of Star Formation in Star-forming Main Sequence Galaxies at z {\ensuremath{\sim}} 2}",
      journal = {\apj},
         year = 2018,
        month = may,
       volume = {859},
       number = {1},
          eid = {56},
        pages = {56},
          doi = {10.3847/1538-4357/aabf8b},
archivePrefix = {arXiv},
       eprint = {1704.00733},
 primaryClass = {astro-ph.GA},
       adsurl = {https://ui.adsabs.harvard.edu/abs/2018ApJ...859...56T}
}

@ARTICLE{nelson2016,
       author = {{Nelson}, Erica June and {van Dokkum}, Pieter G. and {F{\"o}rster Schreiber}, Natascha M. and {Franx}, Marijn and {Brammer}, Gabriel B. and {Momcheva}, Ivelina G. and {Wuyts}, Stijn and {Whitaker}, Katherine E. and {Skelton}, Rosalind E. and {Fumagalli}, Mattia and {Hayward}, Christopher C. and {Kriek}, Mariska and {Labb{\'e}}, Ivo and {Leja}, Joel and {Rix}, Hans-Walter and {Tacconi}, Linda J. and {van der Wel}, Arjen and {van den Bosch}, Frank C. and {Oesch}, Pascal A. and {Dickey}, Claire and {Ulf Lange}, Johannes},
        title = "{Where Stars Form: Inside-out Growth and Coherent Star Formation from HST H{\ensuremath{\alpha}} Maps of 3200 Galaxies across the Main Sequence at 0.7 < z < 1.5}",
      journal = {\apj},
         year = 2016,
        month = sep,
       volume = {828},
       number = {1},
          eid = {27},
        pages = {27},
          doi = {10.3847/0004-637X/828/1/27},
archivePrefix = {arXiv},
       eprint = {1507.03999},
 primaryClass = {astro-ph.GA},
       adsurl = {https://ui.adsabs.harvard.edu/abs/2016ApJ...828...27N}
}

@ARTICLE{Carnall2018,
       author = {{Carnall}, A.~C. and {McLure}, R.~J. and {Dunlop}, J.~S. and {Dav{\'e}}, R.},
        title = "{Inferring the star formation histories of massive quiescent galaxies with BAGPIPES: evidence for multiple quenching mechanisms}",
      journal = {\mnras},
         year = 2018,
        month = nov,
       volume = {480},
       number = {4},
        pages = {4379-4401},
          doi = {10.1093/mnras/sty2169},
archivePrefix = {arXiv},
       eprint = {1712.04452},
 primaryClass = {astro-ph.GA},
       adsurl = {https://ui.adsabs.harvard.edu/abs/2018MNRAS.480.4379C}
}

@ARTICLE{Park2024,
       author = {{Park}, Minjung and {Belli}, Sirio and {Conroy}, Charlie and {Johnson}, Benjamin D. and {Davies}, Rebecca L. and {Leja}, Joel and {Tacchella}, Sandro and {Mendel}, J. Trevor and {Benton}, Chlo{\"e} and {Bugiani}, Letizia and {Emami}, Razieh and {Khoram}, Amir H. and {Li}, Yijia and {Maheson}, Gabriel and {Mathews}, Elijah P. and {Naidu}, Rohan P. and {Nelson}, Erica J. and {Terrazas}, Bryan A. and {Weinberger}, Rainer},
        title = "{Widespread Rapid Quenching at Cosmic Noon Revealed by JWST Deep Spectroscopy}",
      journal = {\apj},
         year = 2024,
        month = nov,
       volume = {976},
       number = {1},
          eid = {72},
        pages = {72},
          doi = {10.3847/1538-4357/ad7e15},
archivePrefix = {arXiv},
       eprint = {2404.17945},
 primaryClass = {astro-ph.GA},
       adsurl = {https://ui.adsabs.harvard.edu/abs/2024ApJ...976...72P}
}

@ARTICLE{spilker2022,
       author = {{Spilker}, Justin S. and {Suess}, Katherine A. and {Setton}, David J. and {Bezanson}, Rachel and {Feldmann}, Robert and {Greene}, Jenny E. and {Kriek}, Mariska and {Lower}, Sidney and {Narayanan}, Desika and {Verrico}, Margaret},
        title = "{Star Formation Suppression by Tidal Removal of Cold Molecular Gas from an Intermediate-redshift Massive Post-starburst Galaxy}",
      journal = {\apjl},
         year = 2022,
        month = sep,
       volume = {936},
       number = {1},
          eid = {L11},
        pages = {L11},
          doi = {10.3847/2041-8213/ac75ea},
archivePrefix = {arXiv},
       eprint = {2208.13917},
 primaryClass = {astro-ph.GA},
       adsurl = {https://ui.adsabs.harvard.edu/abs/2022ApJ...936L..11S}
}

@ARTICLE{deugenio2024,
       author = {{D'Eugenio}, Francesco and {P{\'e}rez-Gonz{\'a}lez}, Pablo G. and {Maiolino}, Roberto and {Scholtz}, Jan and {Perna}, Michele and {Circosta}, Chiara and {{\"U}bler}, Hannah and {Arribas}, Santiago and {B{\"o}ker}, Torsten and {Bunker}, Andrew J. and {Carniani}, Stefano and {Charlot}, Stephane and {Chevallard}, Jacopo and {Cresci}, Giovanni and {Curtis-Lake}, Emma and {Jones}, Gareth C. and {Kumari}, Nimisha and {Lamperti}, Isabella and {Looser}, Tobias J. and {Parlanti}, Eleonora and {Rix}, Hans-Walter and {Robertson}, Brant and {Rodr{\'\i}guez Del Pino}, Bruno and {Tacchella}, Sandro and {Venturi}, Giacomo and {Willott}, Chris J.},
        title = "{A fast-rotator post-starburst galaxy quenched by supermassive black-hole feedback at z = 3}",
      journal = {Nature Astronomy},
         year = 2024,
        month = nov,
       volume = {8},
        pages = {1443-1456},
          doi = {10.1038/s41550-024-02345-1},
archivePrefix = {arXiv},
       eprint = {2308.06317},
 primaryClass = {astro-ph.GA},
       adsurl = {https://ui.adsabs.harvard.edu/abs/2024NatAs...8.1443D}
}

@ARTICLE{chworowsky2024,
       author = {{Chworowsky}, Katherine and {Finkelstein}, Steven L. and {Boylan-Kolchin}, Michael and {McGrath}, Elizabeth J. and {Iyer}, Kartheik G. and {Papovich}, Casey and {Dickinson}, Mark and {Taylor}, Anthony J. and {Yung}, L.~Y. Aaron and {Arrabal Haro}, Pablo and {Bagley}, Micaela B. and {Backhaus}, Bren E. and {Bhatawdekar}, Rachana and {Cheng}, Yingjie and {Cleri}, Nikko J. and {Cole}, Justin W. and {Cooper}, M.~C. and {Costantin}, Luca and {Dekel}, Avishai and {Franco}, Maximilien and {Fujimoto}, Seiji and {Hayward}, Christopher C. and {Holwerda}, Benne W. and {Huertas-Company}, Marc and {Hirschmann}, Michaela and {Hutchison}, Taylor A. and {Koekemoer}, Anton M. and {Larson}, Rebecca L. and {Li}, Zhaozhou and {Long}, Arianna S. and {Lucas}, Ray A. and {Pirzkal}, Nor and {Rodighiero}, Giulia and {Somerville}, Rachel S. and {Vanderhoof}, Brittany N. and {de la Vega}, Alexander and {Wilkins}, Stephen M. and {Yang}, Guang and {Zavala}, Jorge A.},
        title = "{Evidence for a Shallow Evolution in the Volume Densities of Massive Galaxies at z = 4─8 from CEERS}",
      journal = {\aj},
         year = 2024,
        month = sep,
       volume = {168},
       number = {3},
          eid = {113},
        pages = {113},
          doi = {10.3847/1538-3881/ad57c1},
archivePrefix = {arXiv},
       eprint = {2311.14804},
 primaryClass = {astro-ph.GA},
       adsurl = {https://ui.adsabs.harvard.edu/abs/2024AJ....168..113C}
}

@ARTICLE{morishita2022,
       author = {{Morishita}, T. and {Abdurro'uf} and {Hirashita}, H. and {Newman}, A.~B. and {Stiavelli}, M. and {Chiaberge}, M.},
        title = "{Compact Dust Emission in a Gravitationally Lensed Massive Quiescent Galaxy at z = 2.15 Revealed in 130 pc Resolution Observations by the Atacama Large Millimeter/submillimeter Array}",
      journal = {\apj},
         year = 2022,
        month = oct,
       volume = {938},
       number = {2},
          eid = {144},
        pages = {144},
          doi = {10.3847/1538-4357/ac9055},
archivePrefix = {arXiv},
       eprint = {2208.10525},
 primaryClass = {astro-ph.GA},
       adsurl = {https://ui.adsabs.harvard.edu/abs/2022ApJ...938..144M}
}

@ARTICLE{lorenzon2025b,
       author = {{Lorenzon}, G. and {Donevski}, D. and {Man}, A.~W.~S. and {Romano}, M. and {Whitaker}, K.~E. and {Belli}, S. and {Liu}, D. and {Lee}, M.~M. and {Narayanan}, D. and {Long}, A. and {Shivaei}, I. and {Nanni}, A. and {Lisiecki}, K. and {Sawant}, P. and {Rodighiero}, G. and {Damjanov}, I. and {Junais} and {Dav{\'e}}, R. and {Pappalardo}, C. and {Lovell}, C. and {Hamed}, M.},
        title = "{ALMA Reveals Diverse Dust-to-gas Mass Ratios and Quenching Modes in Old Quiescent Galaxies}",
      journal = {\apjl},
         year = 2025,
        month = dec,
       volume = {995},
       number = {2},
          eid = {L63},
        pages = {L63},
          doi = {10.3847/2041-8213/ae226c},
archivePrefix = {arXiv},
       eprint = {2509.10079},
 primaryClass = {astro-ph.GA},
       adsurl = {https://ui.adsabs.harvard.edu/abs/2025ApJ...995L..63L}
}

@ARTICLE{leja2017,
       author = {{Leja}, Joel and {Johnson}, Benjamin D. and {Conroy}, Charlie and {van Dokkum}, Pieter G. and {Byler}, Nell},
        title = "{Deriving Physical Properties from Broadband Photometry with Prospector: Description of the Model and a Demonstration of its Accuracy Using 129 Galaxies in the Local Universe}",
      journal = {\apj},
         year = 2017,
        month = mar,
       volume = {837},
       number = {2},
          eid = {170},
        pages = {170},
          doi = {10.3847/1538-4357/aa5ffe},
archivePrefix = {arXiv},
       eprint = {1609.09073},
 primaryClass = {astro-ph.GA},
       adsurl = {https://ui.adsabs.harvard.edu/abs/2017ApJ...837..170L}
}

@ARTICLE{johnson2021,
       author = {{Johnson}, Benjamin D. and {Leja}, Joel and {Conroy}, Charlie and {Speagle}, Joshua S.},
        title = "{Stellar Population Inference with Prospector}",
      journal = {\apjs},
         year = 2021,
        month = jun,
       volume = {254},
       number = {2},
          eid = {22},
        pages = {22},
          doi = {10.3847/1538-4365/abef67},
archivePrefix = {arXiv},
       eprint = {2012.01426},
 primaryClass = {astro-ph.GA},
       adsurl = {https://ui.adsabs.harvard.edu/abs/2021ApJS..254...22J}
}

@ARTICLE{leja2019,
       author = {{Leja}, Joel and {Carnall}, Adam C. and {Johnson}, Benjamin D. and {Conroy}, Charlie and {Speagle}, Joshua S.},
        title = "{How to Measure Galaxy Star Formation Histories. II. Nonparametric Models}",
      journal = {\apj},
         year = 2019,
        month = may,
       volume = {876},
       number = {1},
          eid = {3},
        pages = {3},
          doi = {10.3847/1538-4357/ab133c},
archivePrefix = {arXiv},
       eprint = {1811.03637},
 primaryClass = {astro-ph.GA},
       adsurl = {https://ui.adsabs.harvard.edu/abs/2019ApJ...876....3L}
}
\bibliographystyle{aasjournal}

\end{document}